# Effect of cold rolling strain on the microstructural evolution in equimolar MoNbTaTiZr refractory complex concentrated alloy: Comprehensive characterization

Andrea Školáková[1,*], Haruka Katayama[2], Pavel Lejček[1], Orsolya Molnárová[1], Sadahiro Tsurekawa[2], Petr Veřtát[1], Jan Duchoň[1], Jaroslav Čech[3], Petr Svora[1], Ondřej Ekrt[1], Jan Pinc[1]

[1]FZU – Institute of Physics of the Czech Academy of Sciences, Na Slovance 1999/2, Prague 8, 182 00, Czech Republic

[2]Department of Materials Science and Engineering, Graduate School of Science and Technology, Kumamoto University, 860-8555, Kumamoto, Japan

[3]Department of Materials, Faculty of Nuclear Sciences and Physical Engineering, Czech Technical University in Prague, Trojanova 13, Prague, 120 00, Czech Republic

*Corresponding author: skolakova@fzu.cz (A. Š.); Tel: +420266052631

**Abstract**

This work presents a pilot study on a strained complex concentrated alloy based on refractory elements: MoNbTaTiZr. Initially, the as-cast and homogenization-annealed conditions were characterized. After casting, the alloy consists of two solid solutions with BCC 1 and BCC 2 crystal structures. Homogenization annealing promotes the growth, ordering, and refinement of the BCC 2 phase. TEM and AES analyses indicate possible Zr segregation at grain boundaries in the as-cast state. In contrast, annealing followed by cooling results in the formation of Ti-Zr-based particles without segregation. Subsequently, the annealed alloy was cold-rolled, and its microstructure was investigated. During rolling, grain fragmentation occurs within the structure. In addition to the two BCC solid solutions, a phase with an FCC crystal structure is identified after rolling. Its composition corresponds to the $Zr_2Ta$ phase, which is a Laves phase of the $A_2B$ type. Rotational relationships, relatively rare in rolled materials with BCC structures, are identified. The texture components found after 10 % rolling deformation are related to that present after 20 % deformation by a 45°⟨110⟩ rotation, and this component is related to that appearing after 30 % deformation by a 20°⟨100⟩ rotation. However, no distinct rolling texture or clear texture development was observed, although some mutual relationships among preferred orientations can be identified. Schmid and Taylor factor maps demonstrate that, despite deformation, the alloy remains capable of further strain accumulation and plastic deformation. Twinning is also observed after rolling, which may be beneficial, as deformation twinning contributes to improved ductility in the alloy.

## 1. Introduction

Refractory metal high-entropy alloys (RHEAs) are a subgroup of a new generation of high-entropy alloys (HEAs), nowadays more commonly referred to as complex concentrated alloys (CCAs), that contain at least four refractory elements, such as Ti, Zr, Hf, V, Nb, Ta, Mo, W, and Cr. RHEAs differ from traditional HEAs/CCAs based on 3d-transition metals in their ability to retain high strength up to 1600 °C [1]. Senkov et al. first produced equimolar NbMoTaW and VNbMoTaW alloys by arc melting [1]. RHEAs generally crystallize in the fully disordered BCC solid solution with dendritic grains and is either single-phase [1-7] or two-phase [8-12]. As was mentioned, they exhibit superior mechanical properties surpassing those of conventional superalloys, particularly at temperatures exceeding 600 °C. The phase stability of BCC phases plays an important role in the application of RHEAs [13]. It was reported that the single-phase BCC solid solution is stable after heat treatment above 1000 °C [14, 15], which is why RHEAs have attracted considerable interest in the research of high-temperature materials with the potential to replace conventional γ-Ni-based superalloys.

The substitution of heavy elements by light ones significantly reduces the density of these RHEAs and enabled to development another group of RHEAs, namely TiTaNbZrMo/Hf alloys [3, 16]. These alloys offer not only potential applications as structural materials at elevated temperatures but also as materials for load-bearing implants [8, 10, 17-19] used to restore the function at load-bearing sites subjected to high mechanical stresses, fatigue, and wear in daily activities. They are being considered to replace superalloys in many high-temperature applications.

RHEAs generally exhibit poor ductility at ambient temperature, especially when they contain Mo. The reason for poor ductility of RHEAs is still unclear and undefined [13]. The problem in the determination of deformation mechanism lies in the high degree of chemical disorder of HEAs coupled with insufficient thermodynamic and kinetic data. Therefore, the research focusing on this problem is in the center of scientific attention. On the contrary, it is known that the alloy containing Hf instead of Mo shows satisfied compressive plasticity at ambient temperature. The TiTaNbZrHf alloy possesses superior compression and tension ductility at ambient temperature allowing them to be cold rolled [20, 21]. For this reason, research interest is primarily focused on RHEAs with Hf [5, 7, 12, 22-26], whereas their Mo-containing counterparts have been studied less extensively. So far, it is known, that the addition of Mo positively influences mechanical properties, especially the addition of Mo could improve high temperature mechanical properties or hot hardness [4].

The percentage quantity of the individual elements also affects structure and subsequent plastic deformation capability. It was found that e.g. the reducing of Ti content leads to the increasing of amount of main dendrite phase enriched by Ta and Mo [11] which results in the high modulus of RHEA. Therefore, RHEA with higher Ti content exhibits lower Young's modulus than that with reduced amount of Ti [11]. On the other hand, the coarser dendrite structure of RHEA containing lower amount of Ti is positively accompanied by enhanced plastic deformation [11]. It is obvious that the changes in chemical composition differed from the equimolar composition significantly affect the resulting structure and mechanical properties, as well. This suggests the possibility of controlling the structure by modifying the chemical composition.

Due to already mentioned extreme brittleness of TiTaNbZrMo alloy, this alloy has not yet been subjected to severe plastic deformation as TiTaNbZrHf alloy possessing a good workability [6, 27]. A few works reported that the main deformation mechanism in TiTaNbZrHf closely relates to the movement of screw dislocations, localization of deformation in bands, the formation of dislocation dipoles, loops, tangles, shear, and kink bands at ambient temperature during compression at different strain [14, 22, 26, 28]. These RHEAs are prepared mostly using arc melting [11], however, study [25] demonstrated that grain refining via cold-rolling simultaneously increased the strength and ductility of TiTaNbZrHf alloy. The observed resulting grain size was 38 µm and tensile strength and elongation was measured to be 958 MPa and 20 %, respectively [25]. However, no studies have been published on the characterization of Ti-Ta-Nb-Zr-Mo-based alloys prepared by methods other than casting.

Due to their brittleness, the straining of RHEAs containing Mo is not widely practiced and investigated. Only a few published works reported about the evolution of microstructure in TiTaNbZrHf [14, 21, 26, 27]. Therefore, the main goal of this work is to observe the evolution of microstructure of MoNbTaTiZr alloy during straining at ambient temperature. To this end, the as-cast microstructure was thoroughly characterized to verify whether any elemental segregation occurs, which could potentially be a cause of the brittleness observed in these alloys. This analysis reveals previously unreported findings, contributing with new knowledge to the field. The present study shows an unexplored topic, which is important for the further research of cold-worked RHEAs.

## 2. Experimental

The MoNbTaTiZr (at. %) RHEA, referred to as CCA in this work, was prepared from metals of 99.9 % purity by vacuum arc melting in water cooled copper crucible. To achieve a homogeneous distribution of the constituent elements, the alloy was remelted 8 times and flipped after each melt. The subsequent annealing was performed in vacuum furnace Xerion at 1000 °C for 168 h. The chemical composition of as-cast alloy was verified using an Ametek EDAX Orbis X-ray Fluorescence (XRF) analyzer. The content of H, N, and O impurities was determined by means of the Bruker G8 GALILEO high-end analyzer, designed for the automated and precise analysis of O/N/H in metallic alloys.

Rectangular plates with a thickness of 3 mm were cut from the as-cast annealed samples by electric discharge machine and subsequently, cold-rolled with a reduction in thickness of 10, 20, 25, and 30 %. The directions RD (rolling direction), ND (normal direction), and TD (transverse direction) are used throughout the article; their orientation with respect to the specimen and the rolling direction is illustrated in **Fig. 1**. All prepared materials were ground by sandpapers P180–P4000 and polished with colloidal suspension Eposil F diluted by 30 % hydrogen peroxide (volume ratio 1:6). The microstructure was examined using an FEI 3D Quanta 3D field-emission-gun DualBeam scanning electron microscope (SEM) equipped with Focused Ion Beam (FIB) lift-out technique and electron back-scattering diffraction (EBSD) system detector TSL/EDAX Hikari and an FEI Tecnai G2 F20 X-TWIN transmission microscope (TEM). Thin foils for TEM were prepared by already mentioned SEM. TEM measurements were performed at accelerating voltage of 200 kV. EBSD mapping was performed with a step size ranging from 0.1 to 1 μm, acceleration voltage of 20 kV and beam current of 32 nA. The EBSD data were processed by an EDAX OIM Analysis 8 software. The phase composition and lattice parameters were determined by X-ray diffraction (XRD) using a PANalytical X'Pert PRO powder diffractometer with a Co anode (K$\lambda$1 = 0.17890 nm, K$\lambda$2 = 0.17929 nm) for the as-cast samples, and a PANalytical Empyrean powder diffractometer with a Cu anode (K$\lambda$1 = 0.15406 nm, K$\lambda$2 = 0.15444 nm) for the cold-rolled samples. Both instruments were configured in Bragg-Brentano geometry.

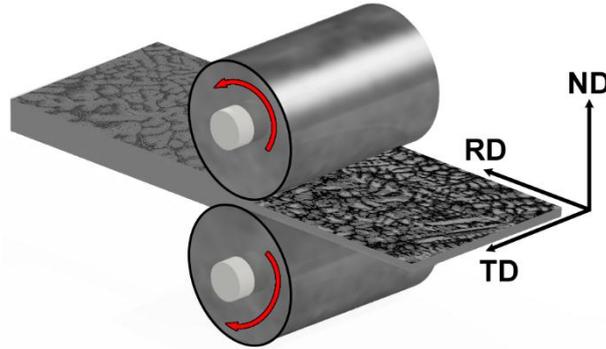

**Fig. 1:** Schematic of directional orientation with indicated directions: RD – rolling direction, ND – normal direction, TD – transversal direction

Auger Electron Spectroscopy (AES) was applied to measure grain boundary chemistry. To realize it, the samples of the dimensions of 3.7 × 3.7 × 16.5 mm$^3$ were fractured at -196 °C by impact bending in the ultra-high vacuum (5×10$^{-8}$ Pa) chamber of an AES facility (ULVAC-PHI, PHI-700). The chemical compositions of both the grain boundary facets and the cleavage surfaces were investigated to characterize the chemical differences between the grain boundaries and bulk material. The parameters of the AES measurements were as follows: primary electron voltage of 10 kV, primary beam current of 10 nA, primary beam diameter of 22 nm (point and line scan mode). To obtain information about the distribution of involved elements in the direction perpendicular to the investigated grain boundary, an area of 2.0 × 2.0 μm$^2$ was analyzed during Ar+ ion sputtering voltage of 1 kV with a sputtering rate of 2.4 nm/min as calibrated by means of a SiO$_2$ reference sample of known thickness.

Nanoindentation tests were conducted to determine the hardness ($H_{IT}$) and indentation elastic modulus ($E_{IT}$) of annealed sample and annealed sample subjected to a 30 % thickness reduction. The measurements were carried out using an Anton Paar NHT$^2$ nanoindentation tester equipped with Berkovich diamond tip. Applied loads were 2 mN for localized measurements and 50 mN for measurements over larger areas. The loading cycle consisted of a 10 s loading phase, a 5 s hold at maximum load, and a 10 s unloading phase. Load-depth data were analyzed using the Oliver-Pharr method [29] according to standard ISO 14577. At least 7 valid indentations were performed in each investigated region and statistically evaluated.

The equilibrium phase diagram was generated using Thermo-Calc software (version 2015a), utilizing the TCHEA4 database for high-entropy alloys.

## 3. Results

### 3.1 Characterization of the initial as-cast and annealed alloy for subsequent cold-rolling

The chemical composition of as-cast alloy was verified by XRF and results are listed in **Table 1** together with the content of impurities. It is evident that the chemical composition of the as-cast alloy was around the required 20 at. %. The impurities such as oxygen and nitrogen were also determined, and measurements confirmed their very low content.

**Table. 1:** XRF chemical composition of the MoNbTaTiZr alloy and the content of impurities

| Element content | | | | | | |
|---|---|---|---|---|---|---|
| at. % | | | | | ppm | |
| Ti | Ta | Zr | Nb | Mo | O | N |
| 19.4 ± 0.2 | 19.3 ± 0.2 | 20.6 ± 0.2 | 20.4 ± 0.1 | 20.3 ± 0.1 | 998.8 ± 73.9 | 555.2 ± 10.7 |

**Fig. 2** shows the X-ray diffraction (XRD) patterns of the non-annealed and annealed as-cast alloys. XRD results revealed the presence of two solid solution phases with BCC crystal structure labelled as BCC 1 and BCC 2 (**Fig. 2**). The XRD patterns did not indicate the formation of any intermetallic compound phase. The BCC 1 phase was identified as major phase in the structure while BCC 2 as minor phase. It is evident that the volume fraction of BCC 2 phase increased (**Fig. 2**) after annealing, but annealing did not lead to the decomposition of BCC solid-solution phases to other phases or the formation of intermetallic compounds. The position of peaks belonging to BCC 2 were shifted to lowers angles and their intensity increased (**Fig. 2**) after annealing, indicating an expansion of the unit cell. The diffraction peaks also became sharper in the annealed alloy. The average lattice parameters were determined to be 0.3361 nm (BCC 1) and 0.3296 nm (BCC 2) for non-annealed alloy while for annealed alloy, the average lattice constants were determined to be 0.3268 nm (BCC 1) and 0.3486 nm (BCC 2). It should be noted that the lattice constant of BCC 2 phase can continuously cover a range of approximately 0.3293 – 0.3297 nm due to the peak asymmetry in the case of non-annealed alloy. The values of lattice constant were compared with those calculated by Vegard's law (**Eq. 1**) resembling the rule of mixtures [30]:

$$a_{mix} = \sum_i c_i a_i,$$  **Eq. 1**

where $c_i$ and $a_i$ are the atomic fraction and lattice constant of the $i$-th element, respectively. The lattice constants of all pure elements are listed in **Table 2**. The calculated lattice constant using Vegard's law should be 0.3322 nm for equimolar MoNbTaTiZr, which is in good agreement with experimental determined lattice constants if we consider that two solid solutions do not have precisely equimolar representation of elements, and there is a redistribution of them between the two solid solutions (BCC 1 and BCC 2).

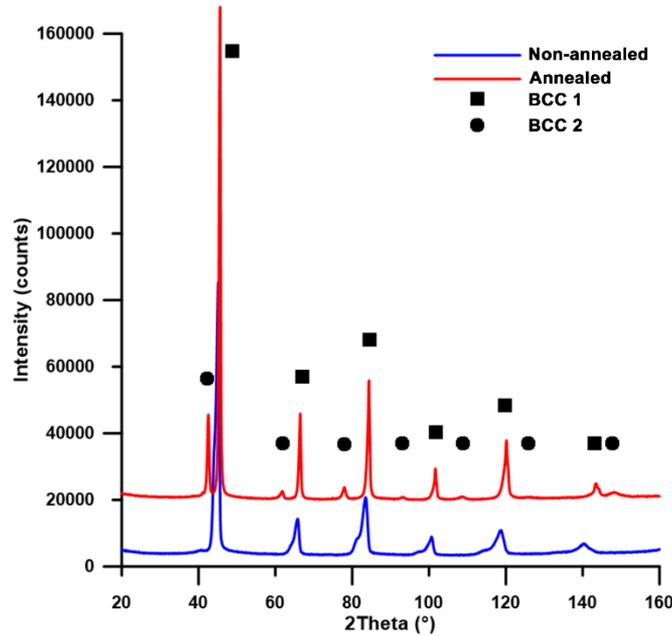

**Fig. 2:** X-ray diffraction patterns for non-annealed and annealed as-cast MoNbTaTiZr alloys

**Table 2:** Lattice constants of the BCC crystal structure of the pure metals and alloy

| Ti [4, 31] | Ta [4, 31] | Zr [4, 31] | Nb [4, 31] | Mo [4] | as-cast alloy (calc.) |
|---|---|---|---|---|---|

| | Lattice constant (nm) | 0.3276* | 0.3303 | 0.3582* | 0.3301 | 0.3147 | 0.3322 |

*The lattice constants of Ti and Zr are extrapolated from their respective high temperature BCC phase to ambient (the crystal structures of Ti and Zr are BCC at high temperatures) using their coefficients of thermal expansion [3]

**Fig. 3** shows the microstructure of non-annealed (**Figs. 3 a, b**) and annealed (**Figs. 3 d, e**) as-cast alloys with corresponding elements distribution maps (**Figs. 3 c, f**) obtained by EDS (Energy Dispersive X-ray Spectroscopy). An equiaxial dendrite microstructure formed during casting and the areas of dendrites differed in the chemical composition due to the redistribution of the individual constituent elements during solidification (**Figs. 3 c, f**). The redistribution of elements led to the formation of two dendrite regions, which could be labelled as the main-dendrite phase exhibiting light-gray contrast at BSE (Backscattered Electrons) image and the minor-dendrite phase with dark-gray contrast at BSE images (**Figs. 3 a, b, d, e**). The minor phase was located in inter-dendrite regions and analysis of chemical composition reveled that this one contained predominantly Nb, Ti, and Zr (**Figs. 3 c, f**), while the main-dendrite regions contained Ta and Mo (**Figs. 3 c, f**). Therefore, Nb, Ti, and Zr had a tendency to enrich the inter-dendrite regions rather than main-dendrite one. This corresponds with the heat of mixing (at melting temperatures) of elements resulting in their redistribution during solidification. The annealing caused the coarsening of the main-dendrite region (**Fig. 3 e**) and low content of Ti was also present in dendrite region (**Fig. 3 f**).

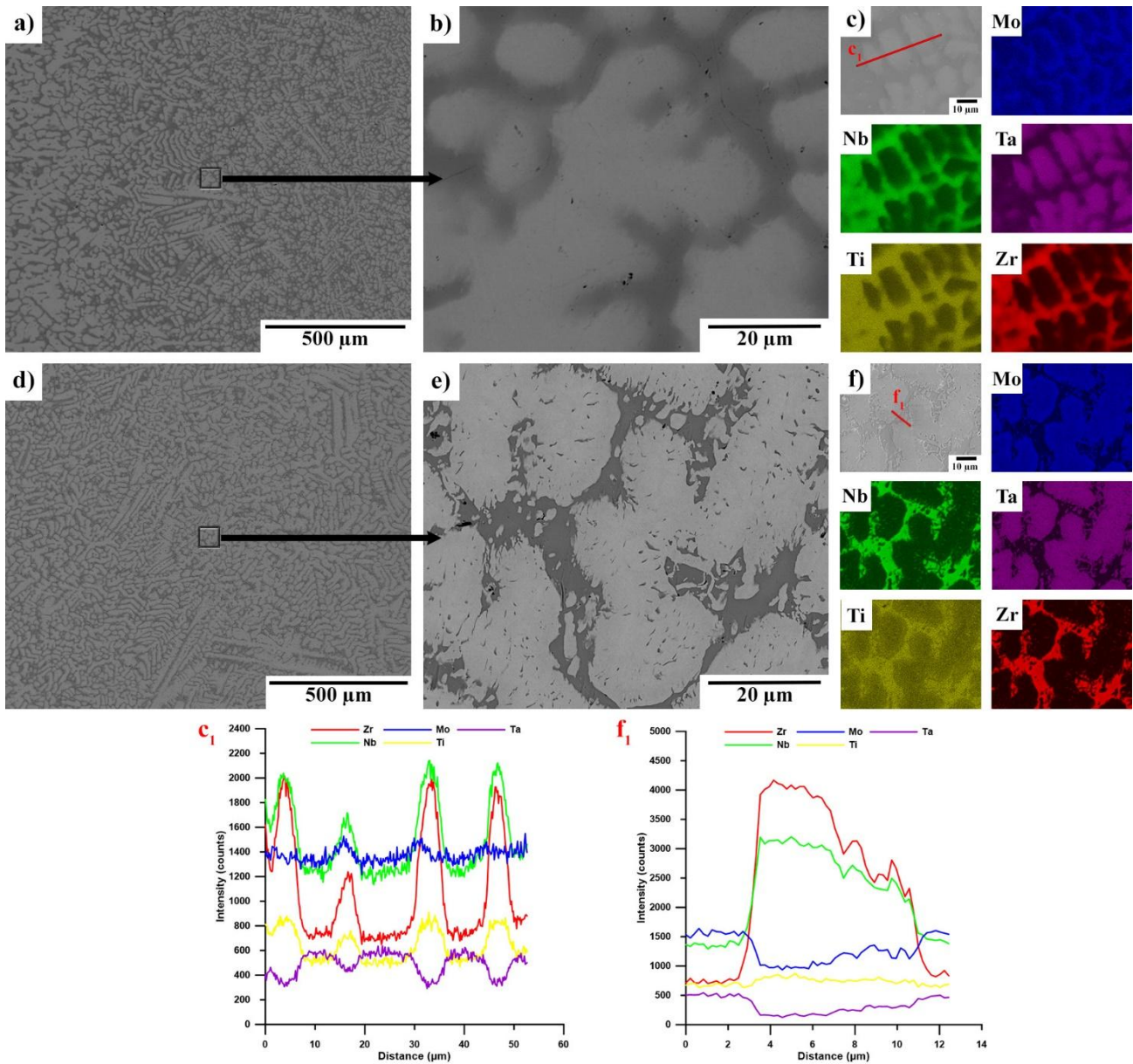

**Fig. 3:** Backscattered SEM images of a) non-annealed as-cast MoNbTaTiZr alloy with b) detail of microstructure, and c) element distribution maps and backscattering SEM images of d) annealed as-cast MoNbTaTiZr alloy with e) detail of microstructure, and f) element distribution maps; EDS line scan showing chemical composition variation for $c_1$) non-annealed and $f_1$) annealed as-cast MoNbTaTiZr alloy

Further analysis of the microstructure, particularly to assess solute segregation in these alloys, was conducted using TEM. The composition changes were found at grain boundaries, dislocations, at boundaries with present particles, or at interfaces of two solid solutions. The results of the TEM analysis for non-annealed as-cast alloy are shown in **Fig. 4**. The bright-field (BF) TEM image along with the corresponding selected area electron diffraction (SAED) pattern (**Figs. 4 a-c**) belong to a region that matched the dark area observed in the BSE image. This corresponds to an interdendritic region enriched in Nb, Ti, and Zr. The second region (**Figs. 4 d-f**) belongs to the bright area observed in the BSE images, indicating a region enriched in Ta and Mo. The chemical composition analysis performed at grain boundaries, solid solution interfaces, and dislocations present in both the bright and dark regions, revealed one specific area indicating Zr segregation within the alloy. This region exhibiting increased Zr content, specifically 34.21 at. % (as determined by STEM-EDS = Scanning Transmission Electron Microscopy-Energy Dispersive X-ray Spectroscopy), was identified at grain boundary (**Fig. 4 g**) with present particle located within the bright region observed in the BSE image. This region is expected to contain higher and comparable amounts of Mo and Ta. However, the quantitative analysis of this small area revealed the following chemical composition (in at. %): Ti 15.43; Zr: 34.21; Nb 16.51; Mo: 20.27, and Ta 13.56 at. %, suggesting the possible segregation of Zr. The profile of Zr concentration is shown as a line scan in **Fig. 4**.

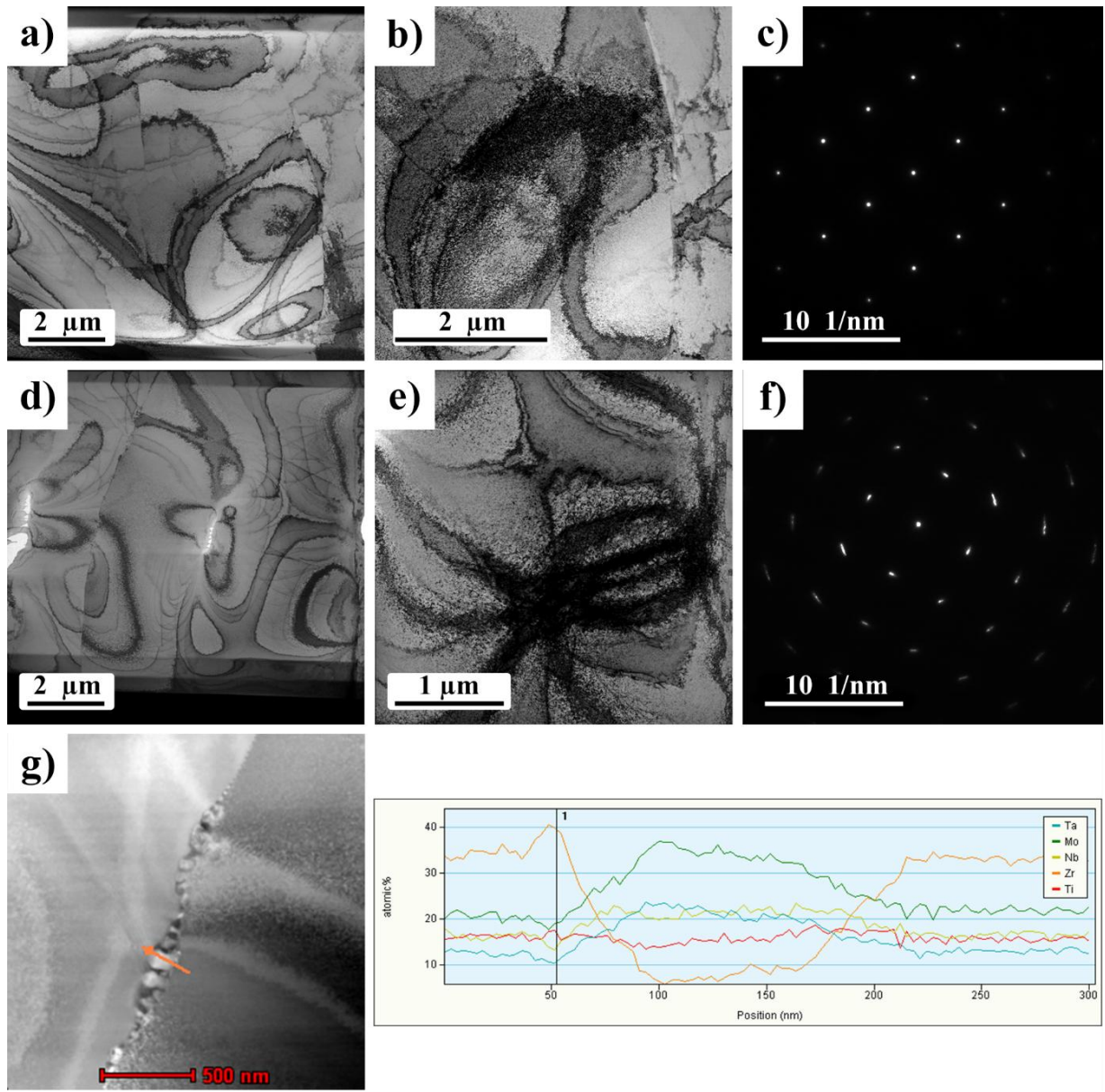

**Fig. 4:** TEM images of non-annealed alloy: a) BF TEM - dark area, b) detailed BF TEM - dark area, c) SAED - dark area, d) BF TEM – bright area, e) detailed BF TEM - bright area, f) SAED - bright area, g) quantified STEM-EDS line profile across particle interface in bright area

**Fig. 5** presents the BF-TEM image of the annealed as-cast alloy, accompanied by the corresponding selected SAED pattern. Dendritic and interdendritic regions are clearly distinguishable, consistent with both the backscattered electron images and elemental distribution maps. Elemental segregation was re-examined in detail. A region intersecting a dislocation is depicted in **Figs. 5 c, d**, with the corresponding line scan analysis shown in **Figs. 5 c, d**. Dislocations were found in both solid solutions, which differ in their chemical compositions. However, no element segregation was observed in either case (**Fig. 5**). Nevertheless, it was found that in the dark region, corresponding in BSE imaging to an area rich in Zr, Nb, and Ti, Mo was also present, whereas Ta was not detected. In contrast, the bright region, enriched in Ta and Mo, contained almost no Zr. It thus appears that annealing did not lead to element segregation, and therefore this condition was selected for subsequent rolling, which will be characterized in the following sections.

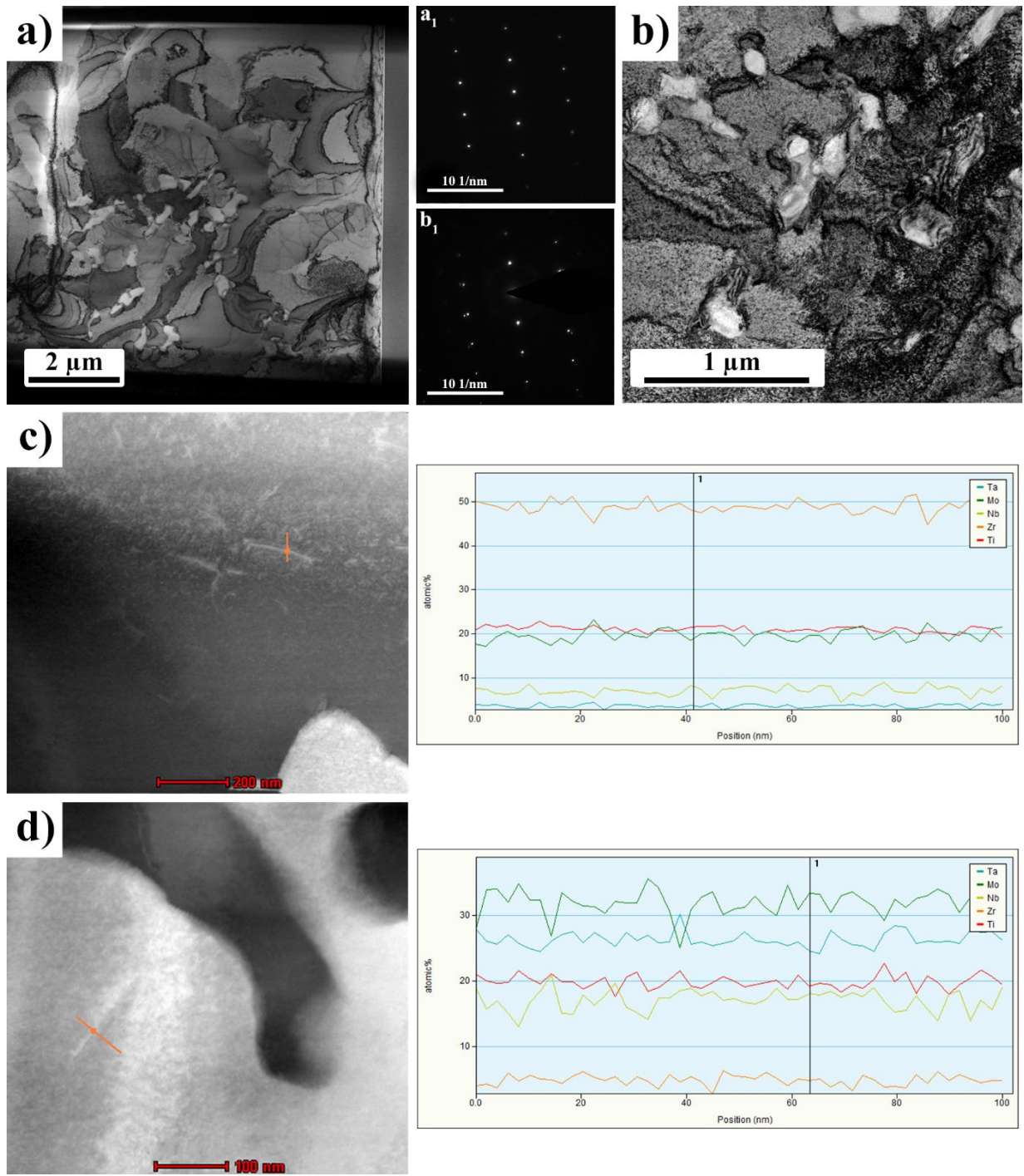

**Fig. 5:** TEM images of annealed alloy: a) BF TEM, a$_1$) SAED from dark area, b) detailed BF, b$_1$) SAED from bright and dark areas, c) quantified STEM-EDS line profile across dislocation in dark area, d) quantified STEM-EDS line profile across dislocation in bright area

**Fig. 6** presents the results obtained from AES analysis. The spectra always show the composition including C and O, as well as the composition excluding these elements. An overview of the fracture surface with the analyzed regions marked is shown in **Fig. 6**, together with the corresponding AES spectra of the non-annealed cast alloy. The marked regions correspond to the locations where grain boundaries, the grain interior, or particles were analyzed: Area 1 (grain boundary, particle 1, particle 2), Area 2 (grain boundary, particle), Area 3 (grain boundary), Area 4 (grain boundary 1, grain boundary 2, grain interior), Area 5 (grain interior 1, grain interior 2). The sample exhibited predominantly brittle fracture, primarily along cleavage planes; however, AES measurements were also performed on several exposed grain boundaries, in some cases also along the fracture surface (see **Fig. 6**, overall view).

Numerous fine particles, each less than 100 nm in diameter, were observed along the grain boundaries, whereas no particles were detected within the grain interiors.

AES measurements were conducted within the regions indicated by the red frames in the magnified SEM image. Quantitative values (**Eq. 2**) were determined using the standardless method [32],

$$X_i = \left(\frac{I_i}{S_i}\right) \bigg/ \sum_{j}^{n} \left(\frac{I_j}{S_j}\right), \qquad \text{Eq. 2}$$

where $I_i$ and $S_i$ are the peak-to-peak height of element *i* and the elemental sensitivity factor of element *i*, respectively. The quantification was done for all (7) elements using the peaks at the AES spectrum. While the presence of O and C is a post-fracturing contamination resulting from the residual atmosphere in the chamber during AES analysis, the quantification was also performed for five metallic elements, only, excluding the contaminants. In the frames in **Figs. 6–10**, the concentrations are given for both ways of quantification. It should be noted that the Mo value may be overestimated due to the overlap of certain Nb sub-peaks with Mo.

Frame 1 in **Fig. 6** shows SEM images and Auger spectra corresponding to Area 1 – the grain boundary. Based on the AES spectra and the reported composition, which was practically equimolar, no elemental segregation was observed at this boundary. The reduction of the quantitative values was caused by contamination by oxygen and carbon. The same result was obtained for another grain boundary studied in this region (Frame 3 in **Fig. 6**). However, in the same analyzed region, an area with a higher concentration of particles was identified (Frames 2 and 4 in **Fig. 6**). The results of the chemical composition analysis revealed that these particles contain a higher concentration of Zr (37.8 at. % and 43.2 at. %). In Area 2 (**Fig. 7 a**), a grain boundary and numerous particles present in high density (Fig. 7 a) were examined. In both cases, an elevated concentration of Zr was observed, approximately 42 at. %. In Area 3, the chemical composition of the grain boundary was analyzed, and the composition corresponded to an equimolar ratio of all constituent elements. Within the analyzed Area 4, the grain interior (**Fig. 7 b**) exhibited a deviated composition, characterized by an increased amount of Zr; however, the concentration was lower compared to the previously observed cases. Grain interiors were also analyzed in Area 5, where both the area with dark contrast and the area with bright contrast were examined. The area with bright contrast contained elements in approximately equimolar composition, with the exception of Ti (**Fig. 7 c**), which was present at a lower concentration (13.4 at. %). In contrast, the area with dark contrast exhibited the highest concentration of Zr, while Ti, Ta, and Nb were present in lower amounts compared to the equimolar composition (see Supplementary material).

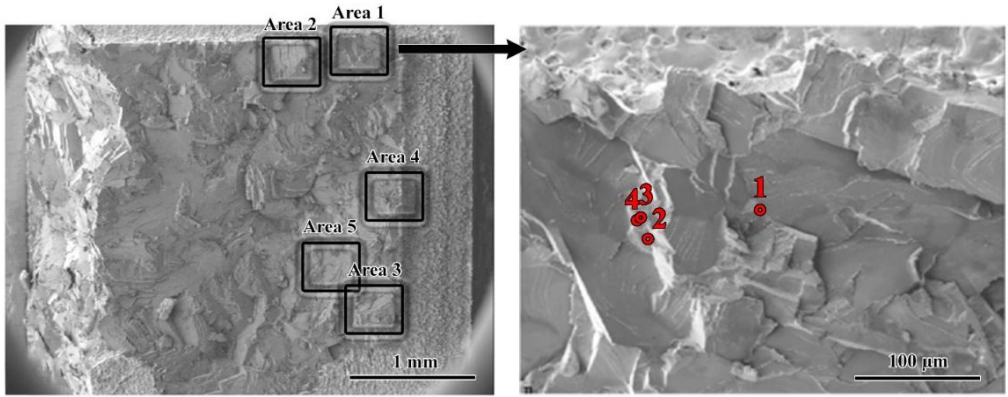
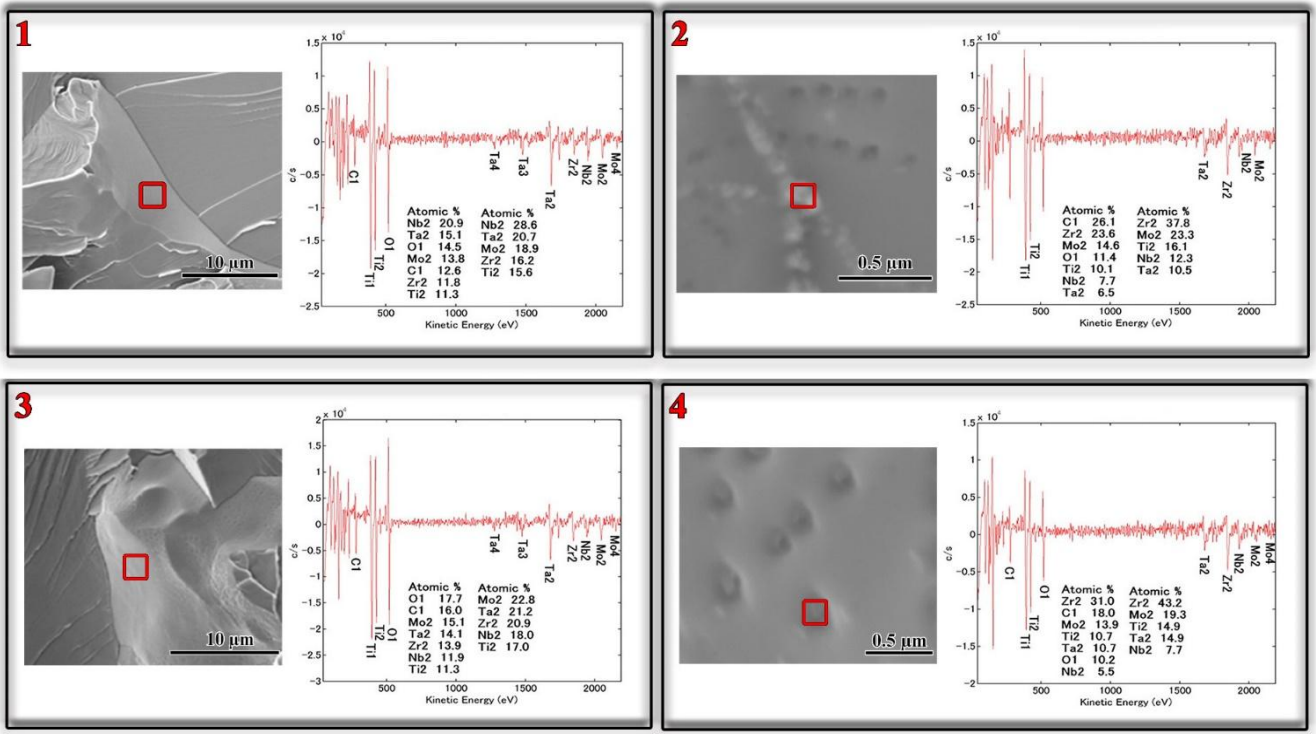

**Fig. 6:** SEM image of fracture surfaces of non-annealed alloy and AES spectra for labeled analyzed areas

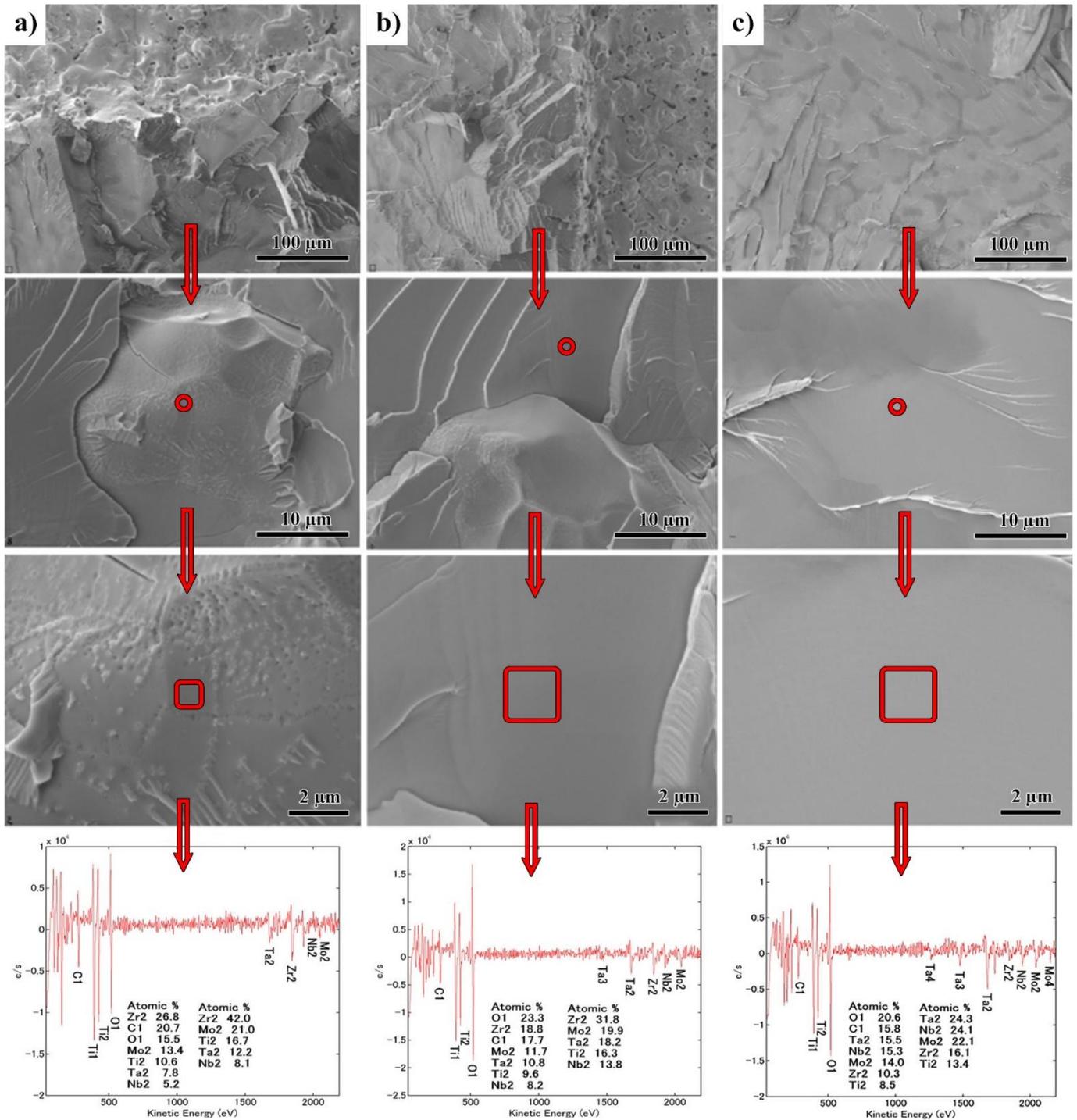

**Fig. 7:** SEM image of fracture surfaces of non-annealed alloy and AES spectra for labeled analyzed areas: a) grain boundary, area 2 from Fig. 6, b) grain boundary, area 4 from Fig. 6, c) grain interior, area 5 from Fig. 6 (the area with bright contrast)

**Figs. 8 a-f** present the results of the AES depth profiling analysis. The analysis was conducted in the vicinity of field of Area 4, which included both transgranular and intragranular fracture surfaces. For measurement convenience, the sample was rotated 90° counterclockwise and tilted by more than 30° prior to sputtering and depth profiling. A decrease in carbon and oxygen concentrations was observed with increasing depth. Although the depth profiling extended to more than five times the displayed depth, no further changes in elemental distribution were detected beyond the surface layer. Therefore, only the data up to approximately 20 nm are shown. **Fig. 8** shows the surface analysis results before and after sputtering at the depth profiling locations. For completeness, spectra for the five principal elements as well as spectra including O and C are presented here although the information about C and O is not shown in the profiles in **Figs 8 a, b**. The depth profiles at the grain boundary (**Figs. 8 b, d, f**) show a gradual decrease in Ti content, which drops below 10 at. %, while the Ta content slowly increases from 20 at. % to 30 at. %. Let us remark that these values do not correspond to the nominal composition of CCA due to presence of numerous precipitates which modify the composition of alloy. In the case of the depth profiles at the grain interior, a gradual decrease in Zr and Ti content is observed with increasing sputter depth (**Figs. 8 a, c, f**). The corresponding AES spectra are shown in **Figs. 8 d, f**, clearly indicating a decrease in Ti content at the grain boundary, as well as a significant increase in Nb content within the grain interior (**Figs. 8 c, e**).

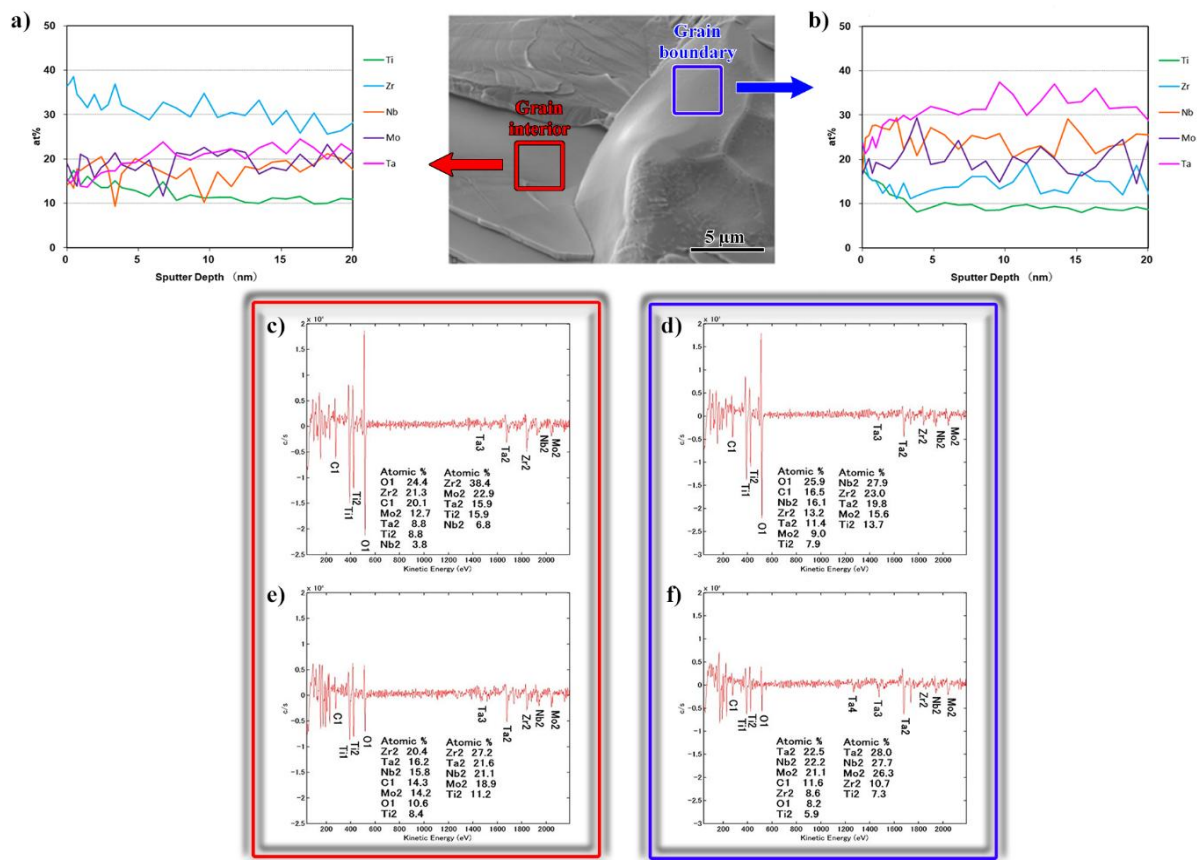

**Fig. 8:** Results of AES depth analysis near area 4 for non-annealed alloy: a) depth profiles at grain interior, b) depth profiles at grain boundary, and measurements c), d) before and e), f) after sputtering at depth analysis points of c), e) grain interior, d), f) grain boundary

The following results relate to the AES analysis of the cast MoNbTaTiZr alloy, which underwent homogenization annealing after casting. For this alloy, five regions on the fracture surface were analyzed. However, the presentation will focus on areas exhibiting more pronounced deviations in chemical composition, since regions such as grain boundaries showed an equal distribution of all five elements, as was observed in the case of the as-cast (non-annealed) alloy. Nearly the entire surface exhibited brittle fracture, with the majority of the fracture surface characterized by cleavage, and some grain boundaries were also observed (**Fig. 9**). It should be again noted that the Mo value may be overestimated due to the overlap of certain Nb sub-peaks with Mo. In Area 1 (**Fig. 9**), the chemical composition of the grain interior was analyzed, revealing that this area contains no Zr and a higher concentration of Nb. The particles found in this region contained a significant amount of Zr accompanied by Ti (**Fig. 9**). In Area 2, a grain boundary was identified that was depleted in Mo and Zr (**Fig. 9**), along with particles rich in Zr and Ti (**Fig. 9**). In Area 3, the grain interior was subsequently analyzed, revealing a very low Ti content. Overall, particles composed mainly of Zr and Ti were observed at the grain boundaries and within the grains.

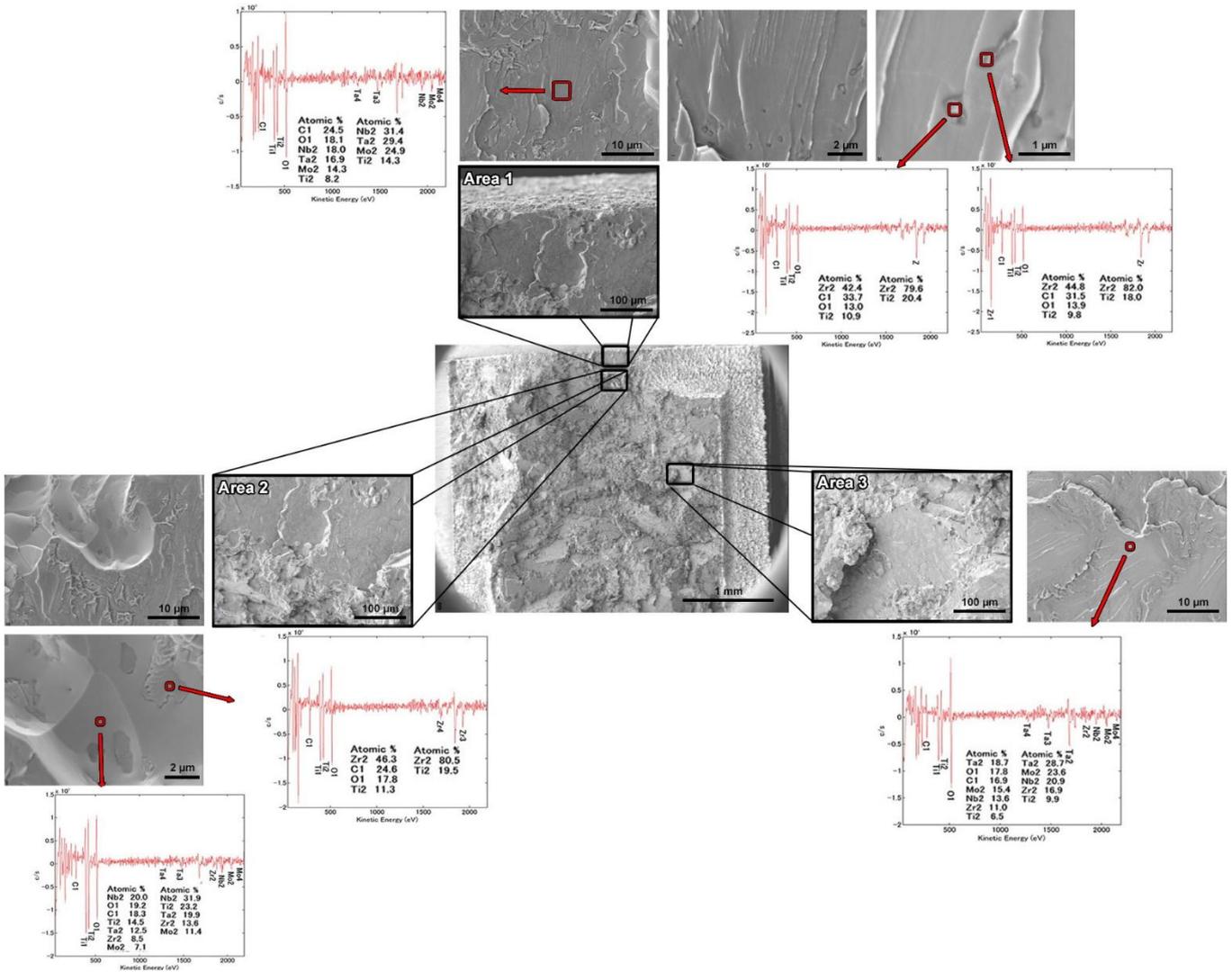

**Fig. 9:** SEM image of fracture surfaces of annealed alloy and AES spectra for analyzed areas

Depth analysis (**Fig. 10**) was also performed in Area 1. Depth analysis showed that contaminants C and O in the surface layer at the grain interior and grain boundary decreased in the depth direction. For the depth profiles of C and O, except at the surface, the data are at background level. As for Zr, it is detected in the surface analysis (**Fig. 10 b**) at the grain boundary before sputtering and within the grain interior after sputtering, but the peak intensity is extremely weak, and the depth profiles are almost at background level. Depth profiles (**Fig. 10**) show a decrease in Ti and Zr concentrations with sputter depth, observed both at the grain boundary (**Figs. 10 b, d, f**) and within the grain interior (**Figs. 10 a, c, e**). Measurements taken before and after sputtering at the depth analysis points reveal that, in the case of the grain boundary, Zr disappears after sputtering, whereas in the grain interior, Zr is present after sputtering despite being absent before sputtering (**Figs. 10 c, d, e, f**).

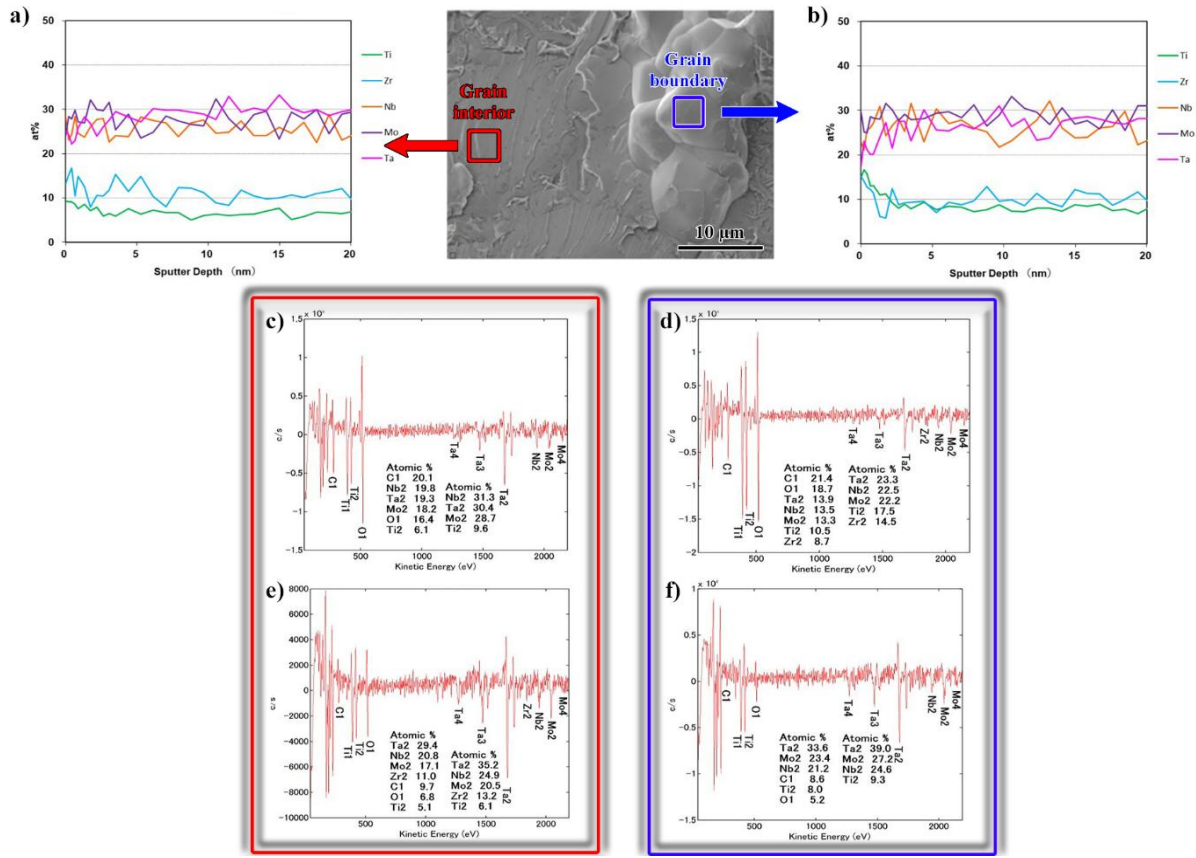

**Fig. 10:** Results of AES depth analysis near area 4 for annealed alloy: a) depth profiles at grain interior, b) depth profiles at grain boundary, and measurements c), d) before and after e), f) sputtering at depth analysis points of c), e) grain interior, d), f) grain boundary

### 3.2 Microstructure evolution during cold rolling

The XRD results of annealed cold-rolled alloys are shown in **Fig. 11**. Phase analysis revealed the presence of two solid solutions with BCC 1 and BCC 2 crystal structures after all rolling paths. Interestingly, a minor contribution from an additional phase that could be indexed as an FCC structure was detected after a 30 % thickness reduction, with its distinct peaks labelled in **Fig. 11**. A comparison of the lattice constants of the BCC 1 and BCC 2 solid solutions before and after annealing and rolling shows no significant change, indicating that the lattice parameters—and consequently the unit cell volumes—remained unaffected by the rolling process (**Table 3**). Therefore, rolling primarily influenced the grain orientation and grain fragmentation rather than altering the crystal structure.

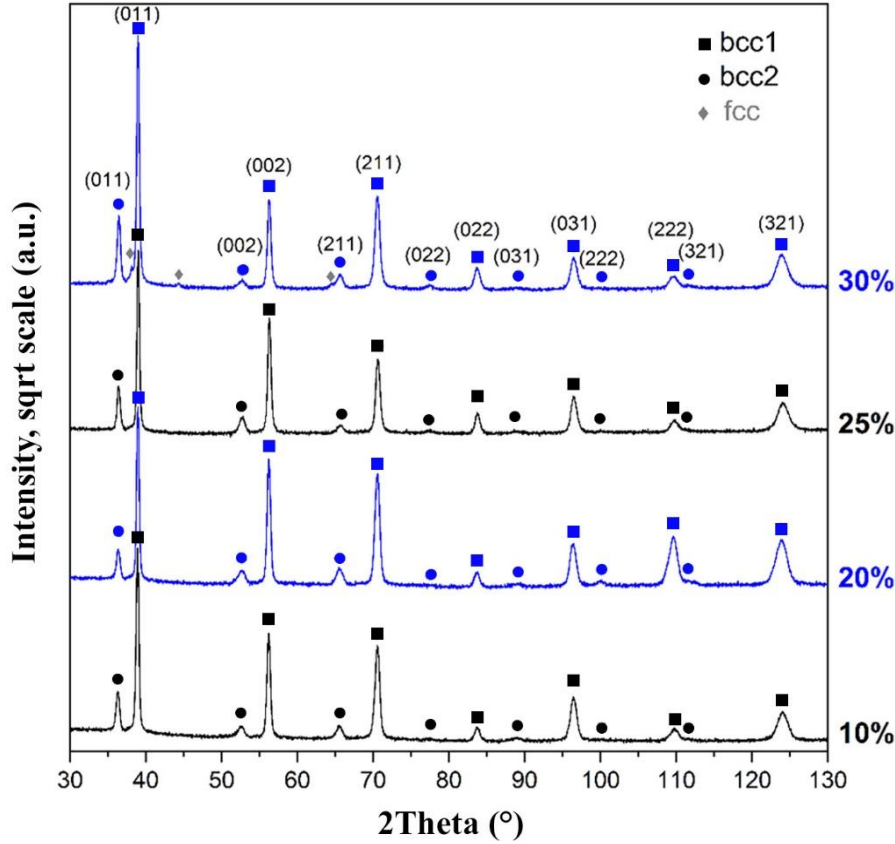

**Fig. 11:** X-ray diffraction patterns for annealed- cold-rolled MoNbTaTiZr alloys

**Table 3:** Lattice constants of the BCC crystal structure in the rolled alloy

| Thickness reduction | 10 % | 20 % | 25 % | 30 % |
|---|---|---|---|---|
| Lattice constant (nm) | BCC 1 = 0.3266 | BCC 1 = 0.3271 | BCC 1 = 0.3267 | BCC 1 = 0.3269 |
|  | BCC 2 = 0.3488 | BCC 2 = 0.3490 | BCC 2 = 0.3487 | BCC 2 = 0.3487 |

For illustration, the microstructure of the annealed cold-rolled alloys with the highest reduction is shown in **Figs. 12 a – c**. The structure after rolling in the annealed state (**Fig. 12 a**) was much more cracked than in the as-cast state (**Fig. 3 d**). Cracks formed mainly perpendicular to the RD. Compared to the as-cast state, the distribution of individual elements within the dendrite and inter-dendrite regions has not changed (**Fig. 12 c**). In **Fig. 12 b**, a higher magnification image of the region outside the crack is shown. As evident, the microstructure consists of three distinct areas differing in chemical composition. A newly observed grey region is present between the dendritic and inter-dendrite regions. The chemical composition of this region likely corresponds to areas where the dendritic and inter-dendrite regions interpenetrate. This region exhibits an increased concentration of Ti (**Fig. 12 c**). From **Fig. 12 c**, it is also apparent that grey region is more damaged, as indicated by the presence of islands. These islands are enriched in Mo and Ta. Tantalum and molybdenum then formed thin bands penetrating the inter-dendrite region (**Fig. 12 c**).

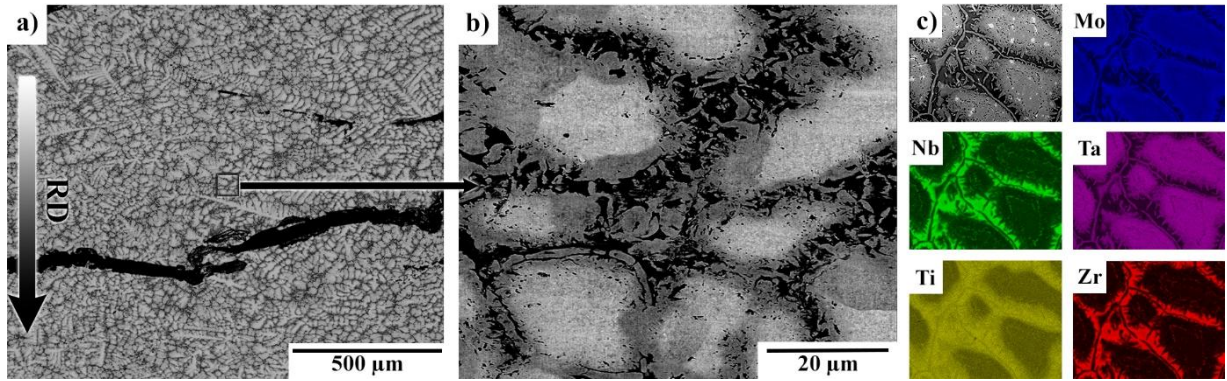

**Fig. 12:** Backscattered SEM images of a) annealed MoNbTaTiZr alloy after 30 % cold-reduction with b) detail of microstructure, and c) element distribution maps

It should be emphasized that all samples exhibited extensive cracking, and the microstructure was analyzed exclusively in regions located between the cracks. **Figs. 13 a – d** show the evolution of the microstructure as the reduction increased during rolling. The RD was from left to right. As can be seen, a 10 % thickness reduction did not significantly affect microstructure. It retains its original and relatively continuous dendritic network, corresponding to as-cast state with a pronounced dendritic structure. Noticeable chemical segregation between the dendritic and interdendritic regions is visible by the bright/dark contrast in BSE mode. The bright dendritic regions are primarily enriched in Ta, Mo, and Ti, whereas inter-dendrite regions (dark) are enriched in Nb, Zr, and partially also in Ti. The sharp and well-defined microstructural boundaries could be observed suggesting only minimal evidence of plastic deformation is present. At a 20 % thickness reduction, the dendritic arms partially fracture into irregular fragments ranging from 30 to 50 µm in size. Their edges are wavy and still follow the Zr/Nb chemical segregation pattern. At the fragment boundaries, the initial formation of distinct regions can be observed. The original dendritic morphology begins to disappear, but it is still relatively retained. An increase in thickness reduction to 25 % caused significant deformation and fragmentation of the original structure. A finer separation becomes apparent, and the original dendrites are no longer distinguishable. The microstructure exhibits a transition from a mosaic- to band-like appearance, without the formation of uniform recrystallized grains. The presence of subgrain structures is likely. The highest thickness reduction achieved during rolling caused an almost complete breakdown of the original as-cast microstructure. Uniform grains with smooth boundaries or typical orientation homogeneity are still not observed, indicating that recrystallization has not occurred. Elemental segregation remains visible (on the base of BSE mode), confirming the absence of significant diffusion, which is extremely slow in these alloys.

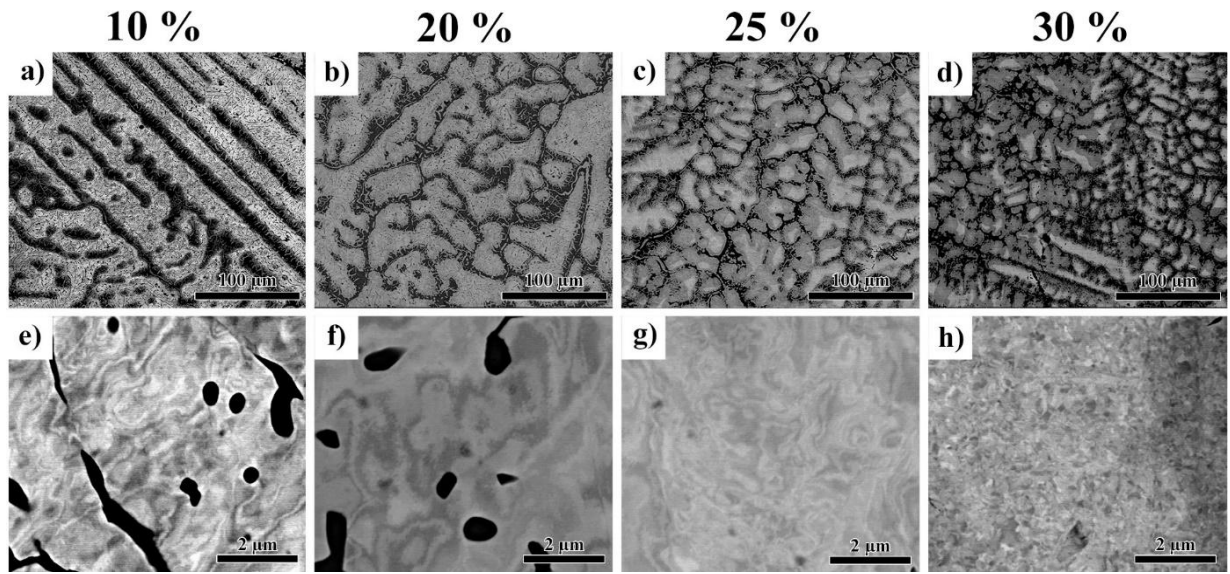

**Fig. 13:** Backscattered SEM images of microstructure evolution during rolling at 10 %, 20 %, 25 %, and 30 % reduction (a–d), with corresponding high-magnification images (e–h) for annealed cold-rolled alloy

A more detailed view of the structure (**Figs. 13 e – h**) reveals the behavior of both present solid BCC solutions. At a 10 % thickness reduction, the secondary (dark) phase forms continuous channels between the primary regions. With increasing deformation, these channels progressively fragment into isolated islands. At 20–25 % thickness reduction, the secondary phase begins to fracture along subgrain boundaries within the primary matrix, leading to a fine-scale dispersion. At approximately 30 % thickness reduction, the dispersion of the secondary phase becomes nearly uniform, creating favorable conditions for subsequent recrystallization. Evidence of polygonization is observed: subgrain boundaries-visible as fine, curved lines-separate regions with slight misorientation, indicating dynamic recovery without full recrystallization. The primary and secondary phases still exhibit considerable continuity, with only moderate elongation retained in the RD. Therefore, as deformation progresses, both phases begin to deform more visibly. Elongation and local waviness (wave-like morphology) become apparent. The secondary dark phase fragments into smaller segments, while the primary matrix develops fine, curved subgrain boundaries. The dispersion of the secondary phase becomes nearly homogeneous, and the primary matrix shows signs of pronounced microstructural changes. A very fine subgrain structure develops, without regular band formation. The secondary phase becomes almost completely dispersed within the primary matrix, resulting in a very fine, nearly homogeneous two-phase mosaic.

The IPF (Inverse pole figure) maps in **Figs. 14 a - d** with appropriate detailed images are presented along the RD for gradual thickness reductions of the alloy. As observed, with increasing thickness reduction, the microstructure becomes more fragmented and contains a higher number of smaller grains. This is further confirmed by the gradually increasing length of high-angle grain boundaries (HAGBs, black lines). Regarding the analysis of low-angle grain boundaries (LAGBs, white lines), their highest number was observed after a 10 % thickness reduction. LAGBs are present both along HAGBs and within the grains themselves. Following such a low degree of thickness reduction, the microstructure is still largely dominated by the original as-cast morphology (**Fig. 14 a**). Subsequently, the frequency of LAGBs (5 – 15°) decreased significantly. With a 20 % thickness reduction, a network of LAGBs appears along the edges of the original grains, indicating the activation of polygonization and the formation of subgrain structures. The microstructure consists of large original dendritic grains with increasing ratio of fragmentation (**Fig. 14 b**). At a 25 % thickness reduction, the length of HAGBs increases further, and the subgrain network (LAGBs) becomes more pronounced. Small recrystallization nuclei begin to appear at the boundaries of highly deformed regions. The formation of small nuclei is observed, accompanied by the development of a granular, mosaic-like microstructure. (**Fig. 14 c**). Finally, at a 30 % thickness reduction, the proportion of HAGBs increases noticeably, which may indicate a progressing structural transformation. The microstructure is heavily fragmented, consisting of a fine-grained mixture of fragmented blocks ready for boundary migration during the potential subsequent annealing (**Fig. 14 d**). The **Fig. 14 a – d** also indicates that deformation occurred locally within the sample rather than throughout the entire specimen. The IPF maps of the initial as-cast states are provided in the Supplementary.

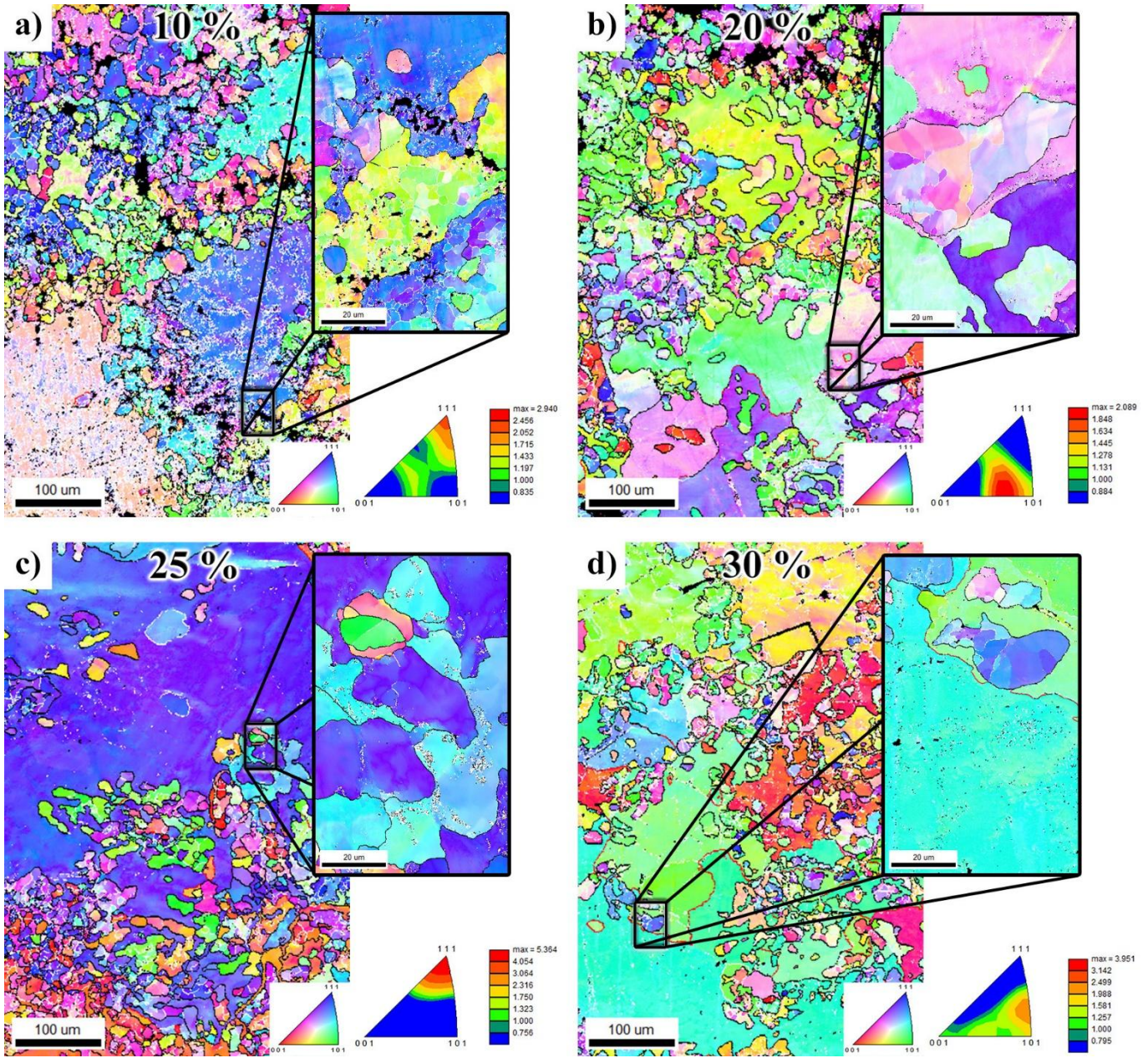

**Fig. 14:** IPF maps (along the RD) showing microstructure evolution during rolling at 10 %, 20 %, 25 %, and 30 % reduction (a–d), with corresponding texture components displayed as inverse pole figure (IPF) triangles along the (RD). The white and black lines represent LAGBs (5 – 15°) and HAGBs (>15°), respectively, and red lines Σ3 CSL boundaries

This development confirms that cold rolling first induces subgrain formation (details in **Figs. 14 a – d**), followed by extensive grain fragmentation through HAGBs and the selective preservation of low-energy CSL interfaces (red lines in IPF maps), resulting in a microstructure that is prepared for efficient recrystallization annealing.

Regarding crystal orientation along the RD, the IPF map for the sample subjected to a 10 % thickness reduction reveals a random crystallographic texture (**Fig. 14 a**), which indicates that no crystallographic direction was preferentially aligned with the RD. After a 20 % thickness reduction, small clusters of green-yellow grains begin to appear, indicating the initial rotation of grains toward [101] and approximately 30° away from the [001] orientations (**Fig. 14 b**), suggesting the early stages of fiber texture formation. However, in our case, the microstructure was not composed of α or γ fibers even after the completion of rolling. The map for sample after a 25 % thickness reduction is predominantly blue to purple (**Fig. 14 c**), indicating the presence of a strong [111] texture.

This suggests that grains are rotating such that their [111] directions align with the RD, as these BCC slip systems exhibit the lowest critical resolved shear stress during deformation. After a 30 % thickness reduction, the IPF map is predominantly green, indicating a prevailing [101] texture (**Fig. 14 d**). With increasing deformation, grains rotate from the [111] orientation toward the more stable [101] direction.

The evolution of crystallographic texture was analyzed in more detail using IPF in the RD (with crystallographic directions mapped in a stereographic triangle bounded by [100]) with color gradients from blue (low intensity) to red (high intensity) indicating the strength of texture components (**Figs. 14 a – d**). At a 10 % thickness reduction (**Fig. 14 a**), the IPF revealed a moderate crystallographic texture, with a dominant intensity peak near the [111] vertex. This suggests that a notable fraction of grains had their [111] crystallographic direction aligned with RD. The color distribution exhibited a broad spread, transitioning from blue in the outer regions through green and yellow to a concentrated red area at [111]. This pattern indicates the initial formation of a {111}⟨110⟩ texture component, characteristic of early deformation stages in BCC materials. A 20 % thickness reduction led to a marked evolution of the crystallographic texture (**Fig. 14 b**), characterized by a shift of the intensity maximum toward the central region of the IPF triangle. This central region corresponds to orientations where RD is approximately equally distant from [001], [101], and [111], potentially indicating the formation of a [125] parallel to RD texture component or a mixture of orientations. A strong red area was near the center of IPF triangle. This configuration deviates from typical BCC rolling textures, which often develop along the α-fiber (⟨110⟩ parallel to RD). The appearance of the central peak may indicate a transient stage, which could be possibly driven by the complex deformation behavior typical of CCAs. Simply, at 20 % reduction, the peak shifts to the center of the IPF triangle, suggesting a mixed orientation, possibly around [112]. At a 25 % reduction (**Fig. 14 c**), the texture became more pronounced, with a sharp intensity peak at [111], indicating a strong alignment of grains with [111] parallel to RD. An intense red region at [111], accompanied by extensive blue areas elsewhere, indicates a high degree of crystallographic orientation alignment. This development points to a well-defined {110}⟨111⟩ or {112}⟨111⟩ texture, consistent with advanced stages of rolling in BCC materials. The sharpness of the peak suggests minimal orientation spread and reflects a highly textured microstructure. A further increase in reduction to 30 % (**Fig. 14 d**) shows that the peak shifts toward [101]. The color distribution reveals a broader dark orange region near [101] and a secondary area of increased intensity (yellow to orange) extending toward [101], and a green spot near [001], which indicates an increased orientation spread. This shift toward [101] suggests the development of a ⟨110⟩ parallel to RD texture, which is typical of BCC rolling textures at higher reductions, where the texture tends to stabilize.

Since no heating occurred at any stage of the alloy formation, only deformation-induced processes were expected, rather than recrystallization. However, with increasing rolling reduction, the formation of new grains was observed. Grain size was evaluated in terms of area fraction. **Fig. 15 a** illustrates that the grain size varied during rolling and reached its minimum after a 30 % reduction. As can be seen, the grain size related to area fraction of present grains was determined to be 6.4 μm with large measurement deviations, indicating the coexistence of both extremely large and very small grains. With further rolling, this value decreased, reaching a final grain size of 4.1 μm. It should be noted that the most extreme data points were omitted from the graph to improve clarity and readability. To determine the fraction of recrystallized grains, the grain orientation spread (GOS) was evaluated. GOS is defined as the average misorientation angle between all measured points (pixels) within a grain and the mean orientation of that grain. Based on the GOS values, grains were classified into three categories: recrystallized grains with GOS < 1.5° (strain-free), grains in transition with GOS between 1.5° and 3° (partially recrystallized), and deformed grains with GOS > 3° (high internal strain). **Fig. 15 b** illustrates how the fraction of recrystallized grains increases with increasing rolling reduction.

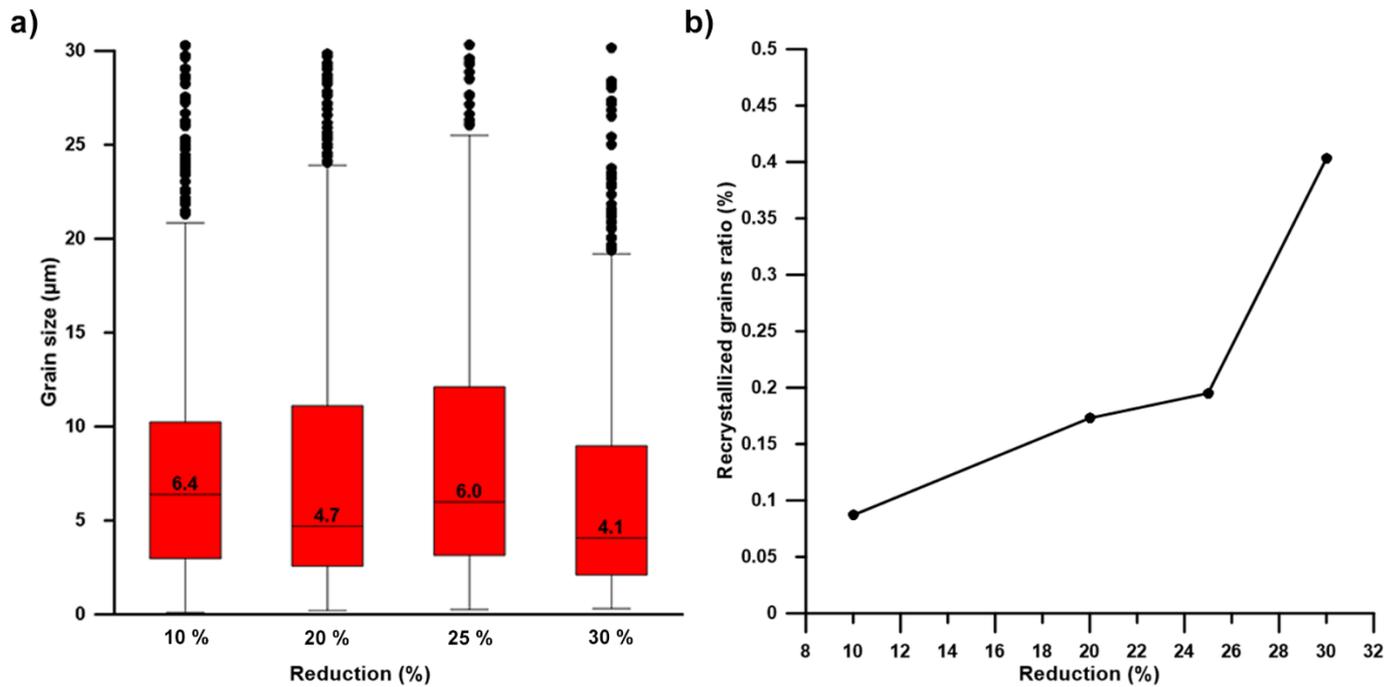

**Fig. 15:** Dependence of a) grain size and b) fraction of recrystallized grains on rolling reduction

The results of the XRD analysis revealed the presence of a phase with a face-centered cubic FCC crystal structure in the sample after 30 % reduction during rolling (**Fig. 11**). Therefore, this sample was further analyzed using TEM in order to characterize this phase. The results are shown in **Figs. 16 a – e**. In addition to confirming that the matrix is composed of a BCC phase (**Fig. 16 c**), particles were observed within the matrix, as shown in the TEM image (**Fig. 16 b**) and the corresponding STEM image of the same region (**Fig. 16 a**). The particle was subjected to detailed analysis. It seems that at least two overlapping particles are present, and both showing internal banding/striation (**Fig. 16 a**). The diffraction patterns indicate a BCC matrix oriented in the ⟨111⟩ direction, along with additional diffraction spots corresponding to the particle (these are the more distant patterns from the matrix, see **Fig. 16 d**). The particles are also oriented in the ⟨111⟩ direction but exhibit a significantly larger lattice parameter than the matrix. However, it is not possible to determine whether the particle has a BCC or FCC crystal structure, since the ⟨111⟩ zone axis yields a similar hexagonal pattern in both BCC and FCC structures. Further electron diffraction analysis along the BCC ⟨110⟩ zone axis (**Fig. 16 e**) confirmed that the particle possesses a BCC crystal structure. The DF (dark field) image (**Fig. 16 b**) obtained from diffraction spot 1 (**Fig. 16 e**) reveals the presence of possible twins within the particle, visible as bands in the DF image.

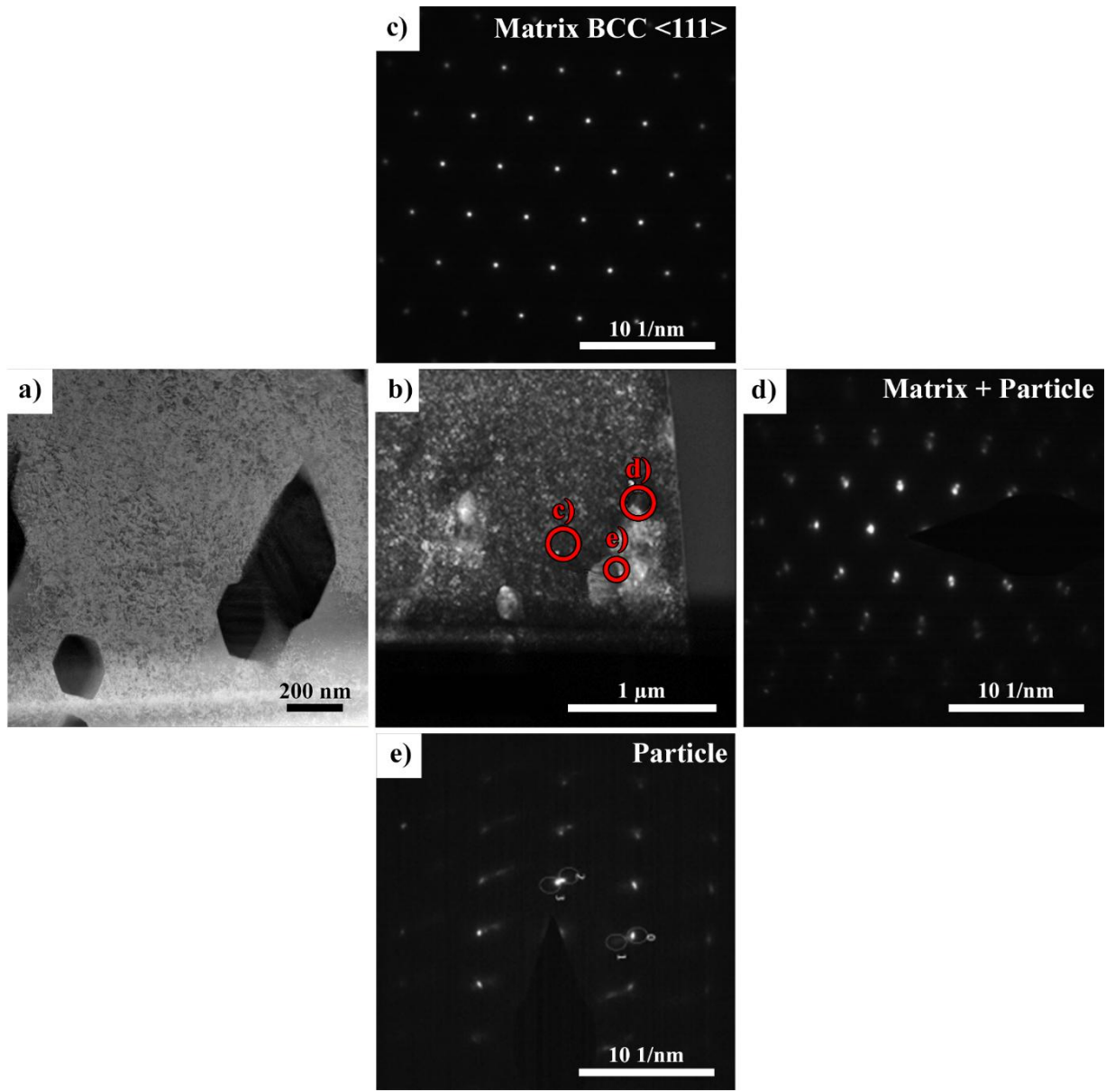

**Fig. 16:** TEM analysis of particle: a) STEM, b) DF, c) SAED from the matrix, d) SAED from the matrix and particle, e) SAED from the particle

Furthermore, the matrix contained particles with an elongated morphology (**Figs. 17 a – d**). This particle exhibited an FCC crystal structure. The diffraction pattern corresponds to the FCC structure along the ⟨110⟩ zone axis (**Fig. 17 c**), and the lattice parameter is a = 0.4668 nm. An EDS analysis of the chemical composition of this particle was performed, and the composition was as follows in at. %: Ti 6.39; Zr 48.65; Nb 14.71, Mo 12.08, and Ta 18.15. However, it should be emphasized that the particle was thin for this analysis, and therefore the measured composition may be influenced by the surrounding matrix. This indicates that the particle is primarily composed of Zr. However, particles that did not exhibit an elongated shape (points 2 and 3 in **Fig. 17**, described above) also showed an elevated Zr content. Specifically, the particle labeled as 2 has the following chemical composition: Ti 15.03; Zr 74.60, Nb 7.35; Mo 1.90, and Ta 1.09 at. %, and the particle labeled as 3 has the following composition: Ti 16.64; Zr 71.40; Nb 7.41; Mo 2.88, and Ta 1.64 at. %. These particles likely correspond to the ones observed during AES analysis (**Fig. 9**).

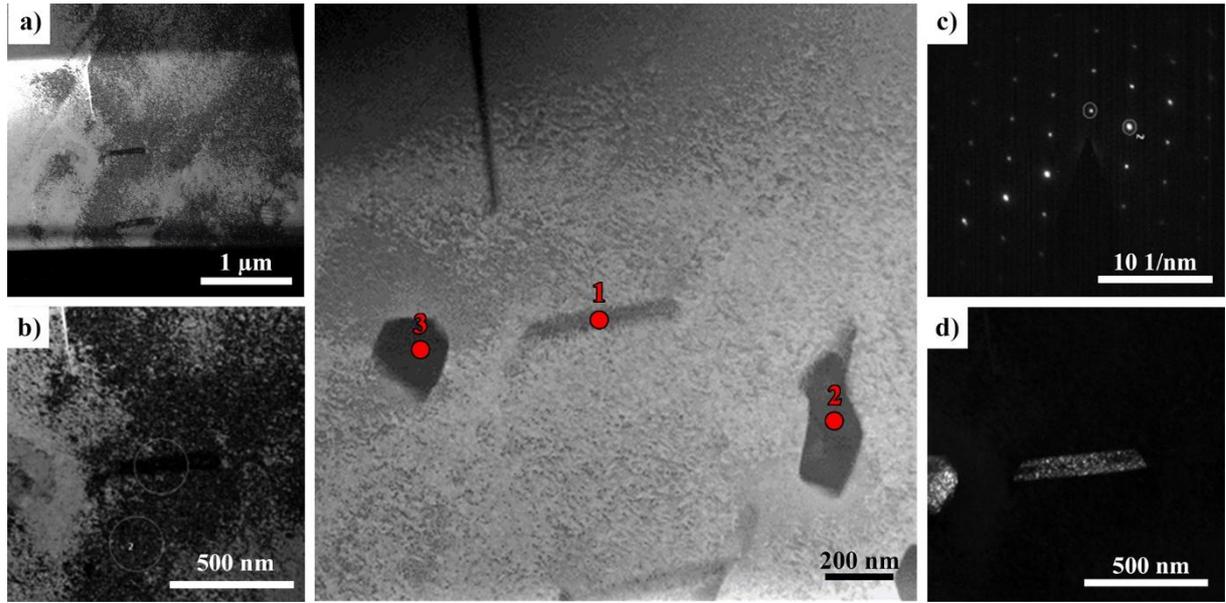

**Fig. 17:** TEM analysis of elongated particles: a) DF of the matrix containing particles, b) high-magnification DF of the matrix with particles, c) SEAD of a particle, d) DF of an individual particle

Due to cracking in the samples, neither stress-strain nor compressive tests could be performed. Therefore, nanoindentation testing was employed to evaluate the mechanical properties. The results are summarized in **Table 4** where the values for bright and dark phases, as identified in the BSE-mode microstructure images, are given. Results clearly indicate that the regions enriched by Ta, Mo, and partially Ti (bright areas) exhibit higher hardness than the Nb and Zr enriched regions (dark areas). Overall, the annealed as-cast alloy exhibited higher hardness compared to the rolled alloy. The highest hardness was measured in the bright regions of the annealed as-cast alloy. This may be attributed to the absence of titanium in these regions, which might otherwise contribute to a reduction in local hardness [33]. Similarly, the Young´s modulus was higher for both the annealed as-cast alloy and the bright regions. A positive outcome is the reduction in the overall Young´s modulus of rolled alloy, specifically its value was: $129.3 \pm 12.9$ GPa. In addition to the bright and dark regions, a grey area was also observed in the rolled alloy. This region exhibited a hardness of $7245 \pm 678$ MPa (at 2 mN load) and a Young's modulus of $107.1 \pm 6.8$ GPa.

**Table 4:** Hardness and Young´s modulus of solid solutions in as-cast and rolled alloys

| MoNbTaTiZr alloy | loading | 2 mN | 2 mN | 50 mN |
|---|---|---|---|---|
| | region | bright | dark | mixed |
| annealed as-cast | hardness H (MPa) | $8391 \pm 530$ | $6608 \pm 545$ | $5790 \pm 516$ |
| | Young´s modulus (GPa) | $163.3 \pm 6.8$ | $117.9 \pm 10.2$ | $137.2 \pm 11.6$ |
| annealed + rolled | hardness H (MPa) | $7387 \pm 375$ | $5724 \pm 298$ | $5707 \pm 845$ |
| | Young´s modulus (GPa) | $163.9 \pm 7.6$ | $107.1 \pm 6.8$ | $129.3 \pm 12.9$ |

## 4. Discussion

This study focused on the characterization of the nearly equimolar MoNbTaTiZr alloy, examining in detail both its as-cast condition and its behavior after cold rolling. Due to its inherently brittle nature, this alloy has not previously been studied in its deformed state. To the authors' knowledge, this work represents the first report in the literature on the cold rolling of a CCAs with the chemical composition MoNbTaTiZr. Nearly all prior investigations of BCC CCAs have demonstrated extremely limited ductility at ambient temperature. Consequently, there is no available literature addressing the deformation-processed form of this specific alloy composition.

**4.1 As-cast and annealed alloy**

The non-annealed and annealed alloy consisted of two solid solutions with a BCC crystal structure (BCC 1 and BCC 2), which is a typical phase composition for this class of alloys confirmed in works e.g. [11, 34]. XRD results (**Fig. 2**) showed lower-intensity peaks corresponding to the BCC 1 in the non-annealed alloy, while their intensities significantly increased after annealing. This can be attributed to insufficient or heterogeneous chemical segregation. Although EDS detects two distinct regions (**Fig. 3 a**), the chemical contrast between them may not be sufficient to produce significantly different lattice parameters (overlapping of peaks). If the two phases have nearly identical lattice parameters, their diffraction peaks may merge or become indistinguishable from those of the dominant phase. Subsequent annealing was performed to homogenize the chemical composition through the chemical segregation across the sample. Although the alloy was repeatedly remelted during casting to promote homogenization, subsequent annealing was conducted based on a previous study [8], which demonstrated improved deformability as a result. Enhanced chemical segregation resulted in a larger difference in lattice parameters, causing the peaks of the BCC 2 to separate more distinctly from those of the BCC 1, and thus become clearly visible. Beside the crystallographic enhancement of the BCC 2, annealing also promoted its growth, resulting in an increased intensity of its diffraction peaks. **Fig. 18** shows diagram obtained via CALPHAD for equimolar alloy MoNbTaTiZr and confirmed the presence of dual phase BCC structure, which corresponds to our results. According to CALPHAD predictions, both BCC phases should be of the B2 type; however, this occurs only under equilibrium conditions with sufficient time and, most importantly, adequate diffusion, which is extremely slow in these alloys. The annealing was performed at 1000 °C, which supported the stabilization of the minor BCC 2 phase and consequently led to the increased intensities in XRD pattern (**Fig. 2**). The formation of BCC 2 phase cannot be attributed to the decomposition of BCC 1 phase due to the presence of a miscibility gap between the BCC phases as was described in work [10]. Moreover, the alloy melt solidifies in a water-cooled copper crucible under a high cooling rate, which can suppress the precipitation of secondary phases during the rapid solidification process [35]. Only elemental segregation during solidification is responsible for the formation of dual-phase structure. The diagram (**Fig. 18**) further shows that during cooling, a phase with an HCP crystal structure may form within the microstructure. However, no such phase was identified by any of the analyses. This can be explained by the fact that the HCP phase could not have formed during slow cooling, suggesting that it is either metastable (as the phase diagram below 400 °C is not known), or that the formation of the HCP phase would require the alloy to be held at the given temperature for a longer period, due to the slow kinetics of this phase's formation.

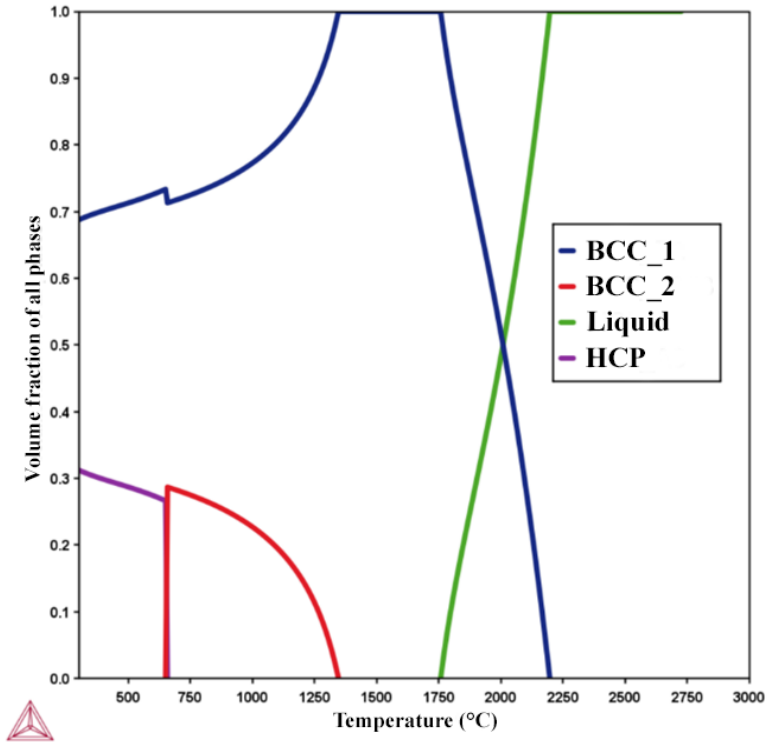

**Fig. 18:** Equilibrium phase diagrams for equimolar MoNbTaTiZr alloy

Due to the increasing reactivity of Ti with temperature, it is not possible to determine the melting point of this alloy using conventional calorimetry. There is a risk of reaction with the crucibles in the calorimeter, and most calorimeters also have temperature limitations. In the referenced study [10], no melting was observed at temperatures below 1773 K, which is consistent with the phase diagram presented here, according to which melting begins at approximately 1750 °C (about 2000 K). Solidus temperature $T_S$ was determined to be 1757 °C (2030 K) and the phase decomposition starts according to phase diagram (**Fig. 18**) at $T_D = 1342$ °C (1615 K). According to an empirical guideline, a $(T_S - T_D)/T_S$ ratio exceeding 0.3 is typically associated with the formation of a single-phase solid solution [35]. With a $(T_S - T_D)/T_S$ ratio of approximately 0.204, the MoNbTaTiZr alloy is thermodynamically inclined to form a dual-phase solid solution upon solidification. Therefore, it is anticipated to remain as a dual-phase metastable solid solution at ambient temperature. This simple empirical rule therefore supports the dual-phase structure of the MoNbTaTiZr alloy, despite the fact that, for example, the TiHfVNbTa alloy exhibits a very similar phase diagram [35]. However, its solidus and decomposition temperatures are significantly lower than those of MoNbTaTiZr, which enables the formation of a single-phase solid solution.

The cast structure was dendritic, which is not surprising given that the alloy is composed of elements with significantly different melting points. Since Ta and Mo possess the highest melting points, the solid solution(s) on basis of these elements were the first to solidify, forming the primary dendrites (**Figs. 3 a, b**). Zr and Ti solidify last subsequently form the interdendritic regions, resulting in chemical segregation between the dendrites. Although Nb has a slightly lower melting point than Mo, its strong affinity for Zr causes it to be concentrated in the interdendritic regions. Homogenization annealing subsequently facilitated a more uniform distribution of Ti within the alloy. As a result, Ti was present in both the interdendritic and dendritic regions (**Fig. 3 b**), which was associated with its segregation in the interdendrite regions [34]. This is consistent with previous works on refractory complex concentrated alloys [8, 11, 36]. The phase composition in both types of alloys revealed the presence of solid solutions with a BCC crystal structure. For such compositions, the heat of mixing is slightly negative or nearly zero, which enables high solubility and consequently promotes the facile formation of solid solutions. The heat of mixing also accounts for the distribution of elements within the dendritic and interdendritic regions, as described in the study [11]. As a consequence of the compositional differences between the dendritic and interdendritic regions, a slight variation in lattice constant arises between these areas, which was confirmed both experimentally (XRD) and through calculations based on Vegard's law.

TEM analysis of non-annealed alloy (**Fig. 4**) revealed that the Ta- and Mo-rich solid solution exhibits an increased concentration of Zr at the grain boundaries compared to other elements, suggesting possible Zr segregation during solidification (cooling). These grain boundaries possess lower atomic cohesion and higher free energy compared to the crystal core. Moreover, for atoms with a larger atomic radius such as Zr = 160 pm (for completeness: Ti 147 pm, Nb 146 pm, Ta 146 pm, and Mo 145 pm), it is thermodynamically favorable to segregate at these boundaries, thereby reducing the overall grain boundary energy [37]. Zr (and Ti) segregation also reduces the cohesion of the grain boundaries [38] which reflects in increased ability of CCA to a brittle intergranular fracture. Zr segregation is also evidenced by AES spectra (**Fig. 6**), which revealed regions near the fracture surface consistently exhibiting an increased Zr concentration in close proximity to particle boundaries. These results correspond to the depth profiles measured at the grain boundary (**Fig. 8 b**), which showed a gradual decrease in Zr concentration with sputtering, while the concentrations of Ta and Mo increased. AES is surface-sensitive, analyzing only the top 1–5 nm of the sample; however, when combined with sequential sputtering, each ion dose removes several atomic layers. Starting from the grain boundary surface, the initial AES measurement detects an elevated Zr concentration (consistent with the segregation observed by TEM). As sputtering progresses, the underlying bulk material (poorer in Zr) is gradually revealed, especially since the main volume of the grain boundary enriched in Zr is confined to the very surface layer. Consequently, the Zr signal measured by AES decreases after several sputtering cycles. TEM analysis did not reveal any elemental segregation in the annealed state of the cast alloy. However, AES analysis identified particles with a high Zr and Ti content at the fracture surface, as well as Ti-rich grain interior regions in which no Zr was detected. The particles must have formed during annealing. Ti and Zr belong to the same group (IV B) of the periodic table and exhibit very similar physical-chemical properties, resulting in a high mutual affinity. The differences in electronegativity and atomic radius are relatively small, enabling Ti and Zr to readily accommodate each other's atoms within their crystal lattices. Since the phase diagram (**Fig. 18**) indicates a continuous α solid-solution region with an HCP crystal structure, the precipitated phases observed are likely part of the same HCP α phase [39]. Subsequent cooling from the homogenization annealing temperature thus led to the formation of this phase in the alloy.

### 4.2 Cold-rolled alloy

Diffraction results revealed that, in addition to two phases with a BCC crystal structure, a minor contribution from a phase with an FCC crystal structure is also present (**Fig. 11**). This phase likely formed during the rolling process. Previous analyses of both the as-cast and less extensively rolled states did not detect it. Its presence was confirmed only in the sample that underwent the highest degree of rolling. The sample was subjected to detailed TEM analysis, which revealed that this phase was present in the matrix in the form of rod-like precipitates (**Fig. 19**). The chemical composition of this phase region is provided in Table listed in **Fig. 19**. As can be seen, the analysis of the chemical composition revealed a predominant presence of Zr and Ta. Neither Zr nor Ta exhibit an FCC crystal structure. Although no FCC phase region exists in the binary Zr–Ta phase diagram [40], nor in the ternary Ti–Ta–Zr system [41], the local formation of an FCC phase within the matrix can occur in complex concentrated alloys. The observed phase is most likely the cubic Laves phase C15 ($MgCu_2$-type), which exhibits an FCC-type diffraction pattern and a composition ratio close to $A_2B$ [42, 43], specifically $Zr_2Ta$, which is consistent with the experimentally determined composition. However, after annealing, two BCC solid solutions were present, in which all elements were dispersed on the atomic scale. This was not a metastable supersaturated state, but a thermodynamically stable solution at the annealing temperature. Under conditions of severe plastic deformation, a possible extremely high density of dislocations and subgrains could generate. These dislocations act as fast diffusion pathways (pipe diffusion) for Zr. Moreover, local shear bands may undergo adiabatic short-term heating [44], which can reach several hundred degrees Celsius, significantly lowering the diffusion barriers for the nucleation and growth of Laves phases. The mechanical energy input and localized heating along dislocation lines simulate these conditions, thereby promoting the in-situ formation of Laves phases during rolling, without the need for subsequent annealing. Furthermore, while the FCC phase is not commonly observed in the binary Zr–Ta system, the multicomponent CCA matrix may facilitate local electronic stabilization of the FCC lattice, particularly in regions enriched with dislocations and lattice defects. The presence of this Laves phase can significantly impair ductility, which may lead to cracking of the alloy during straining [45].

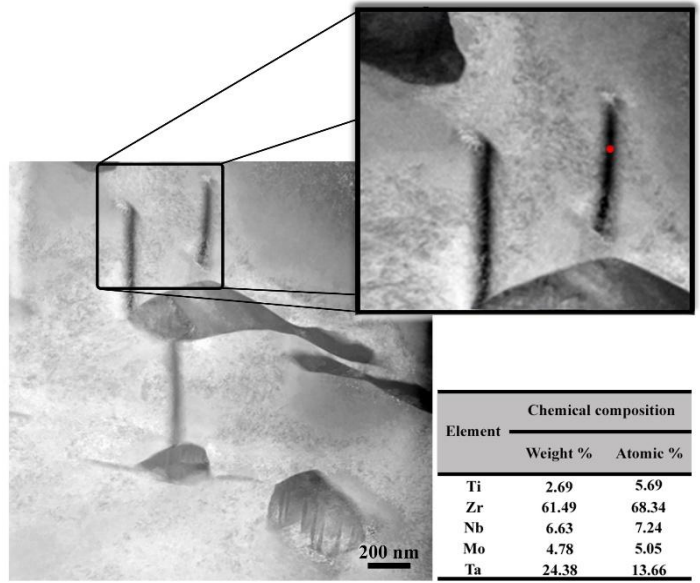

**Fig. 19:** STEM area mapping with a detailed image of rod-like particles and their chemical composition determined by EDS, as presented in the table

To confirm the increased dislocation density in the microstructure, GND (geometrically necessary dislocation) and KAM (Kernel average misorientation) maps are presented (**Fig. 20**). Both maps clearly indicate dislocation accumulation and substructure formation. The KAM maps show increased local misorientation, supporting this accumulation. Dislocations were likely accumulated at subgrain boundaries, which correspond to the green lines on the KAM map (**Fig. 20 b**). This corresponds to the GND map (**Fig. 20 d**), which illustrate the distribution of dislocations arising from the plastic deformation of the material, with their density being directly related to the level of strain and deformation. These maps are primarily used to quantify local deformation, particularly its plastic component. Grains exhibiting low GND density (blue) indicate minimal plastic deformation or recrystallized regions, whereas grains containing internal structures (green lines) likely correspond to dislocation arrangements or subgrains. In contrast, KAM maps represent local plastic deformation (confirmed by Schmid map shown below), internal stresses, and the formation of substructures.

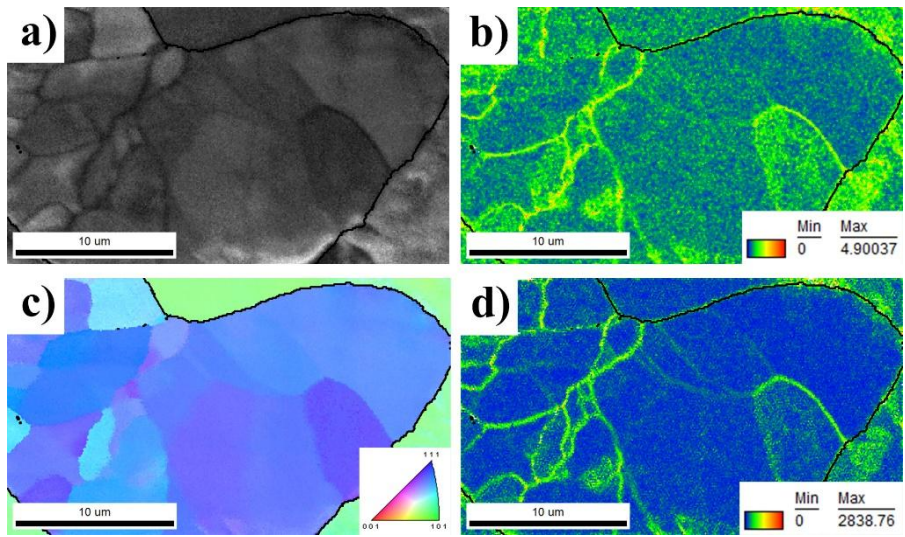

**Fig. 20:** EBSD results for the annealed cold-rolled (30 % reduction) MoNbTaTiZr alloy: a) Image quality (IQ) map, b) KAM, c) IPF map, d) GND map (maximum value × $10^{12}$), black lines represent HAGBs (>15°)

Rolling induced grain fragmentation within the microstructure. Although a high degree of rolling strain was applied to the material, regions that remained almost intact may still exist within the structure, as demonstrated in previous work [45]. The observed texture evolution in MoNbTaTiZr during cold rolling exhibits both similarities and differences compared to conventional BCC metals. The typical α and γ fiber textures [46] did not form due to the presence of two solid solution phases with a BCC crystal structure. The development of such textures generally requires a single-phase alloy, which provides a homogeneous environment for the accumulation of plastic deformation along directions that promote fiber texture evolution, while also minimizing cracking and grain fragmentation. In our case, the alloy exhibited cracking even at low rolling reductions, which can be attributed to the differing deformability of the constituent phases - softer phases deform more readily than harder ones during rolling. This mismatch induces sharp local stress gradients at the phase boundaries, leading to microcrack formation and grain fragmentation. It is evident that the cracking is further promoted by the presence of the previously described particles. The distinct texture evolution observed in CCAs is primarily attributed to their high configurational entropy, which stabilizes randomly ordered solid solutions. This high mixing entropy slows down atomic diffusion, thereby hindering the formation and growth of recrystallization nuclei in preferred orientations after deformation. Sluggish diffusion impedes grain growth along energetically favorable orientations (e.g., α-fiber, γ-fiber). Another significant factor is lattice distortion: differences in atomic radii induce local stress fields, which unevenly affect slip system activation across grains. This results in inhomogeneous slip and disrupts the continuous development of fiber textures. Lattice distortion also increases the activation volume required for dislocation motion, which can modify the relative activity of slip systems, promote complex dislocation interactions, or lead to the activation of non-conventional slip systems in CCAs. This often promotes enhanced twinning activity, as observed in TEM studies, or activates deformation mechanisms that are not conducive to the formation of classical α- and γ-fiber textures. A strong texture in the ⟨111⟩ direction was also observed in the study [27]; however, it was oriented parallel to the normal direction (ND) and a dominant ⟨111⟩ fiber texture was at high reductions. The broadening at 30 % reduction (**Fig. 14 d**) could be attributed to increased strain, leading to more homogeneous deformation or the activation of additional deformation modes, as suggested by studies on similar alloys [45]. These observations align with studies on BCC alloys, where texture evolution during cold rolling involves the development and transformation of specific texture components, influenced by factors such as strain level, slip system activation, and material-specific properties.

   As mentioned above, cold rolling of a material usually results in the formation of more or less pronounced preferential orientations. A detailed study of complex pole figures representation (**Fig. 21**) completes the results based on IPFs summarized in Part 3.2 (**Fig. 14**). The distinct orientations in CCA after particular cold-rolled states do not provide a straight development of unambiguous texture. The situation is also likely complicated by the fact that the material consists of two BCC phases (**Fig. 12** and **Table 3**) and the contributions of individual phases cannot be distinguished in PFs (Pole Figures). In any case, there is no apparent development of the orientations in both the rolling direction and the normal direction. Roughly said, the orientations of the grains in the rolling direction are mainly concentrated in the middle of the orientation triangle (IPF) or at ⟨210⟩ orientation while orientations of the sheet plane are located in the belt between ⟨211⟩ and ⟨311⟩ orientations as well as close to ⟨221⟩ orientation although some other minor components appear at all rolling levels in both cases.

   Although there is no clear rolling texture and its development, some mutual relationships can be found among the preferred orientations. For example, one of the components found after 10 % rolling deformation is related to that occurring after 20 % deformation by rotation of 45°⟨110⟩, and the latter one is related to that appearing after 30 % deformation by rotation of 20°⟨100⟩. However, these rotational relationships are rather rare in the case of the rolling materials with BCC structures. For completeness, the pole figures of the initial condition are provided in the Supplementary.

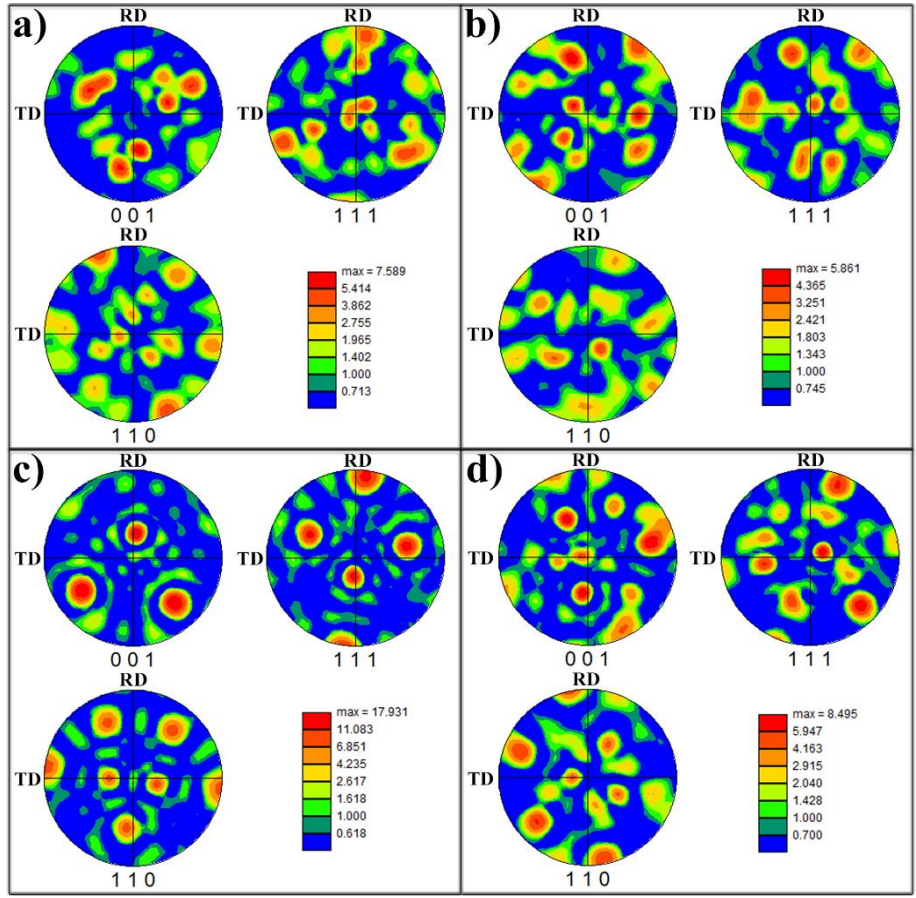

**Fig. 21:** Textures shown by PFs for annealed cold-rolled MoNbTaTiZr alloy at rolling reduction: a) 10 %, b) 20 %, c) 25 %, d) 30 %

The mechanical implications of the observed texture evolution are significant. At a 25 % reduction, the development of a sharp [111] oriented texture is likely to enhance strength due to the alignment of close-packed planes along the loading direction (dislocations have less readily accessible slip planes in this direction); however, this may concurrently reduce ductility as a result of increased crystallographic anisotropy. This sample also showed the most extensive cracking. At 30 % reduction (**Fig. 14 d**), the texture broadens with a noticeable spread towards the [101] direction. This diversification of grain orientations reduces anisotropy and allows more slip systems to activate, leading to a slight decrease in yield strength but a possible corresponding increase in ductility. The observed decrease in texture index with increasing cold reduction may result from a combination of changes in texture components, enhanced grain fragmentation, and transformations occurring during recrystallization. These factors collectively influence the overall material texture and can account for the observed reduction in texture index at higher degrees of deformation (**Fig. 14**). Greater cold reduction promotes significant grain fragmentation and an increase in dislocation density, both of which contribute to orientation dispersion and a corresponding decrease in the texture index. An increase in the fraction of recrystallized grains was also observed (**Fig. 15 b**) at higher levels of reduction. Nucleation and grain growth during recrystallization influence the resulting texture index. Following deformation, recrystallization may occur, during which new grains form and grow. If these newly formed grains are randomly oriented or adopt orientations that do not reinforce the original texture, a decrease in the texture index may result.

The authors were aware of the challenges associated with forming this CCA, particularly given the absence of any published literature on the subject. Preliminary results, however, indicated that these alloys possess sufficient ductility. For this reason, we performed cold rolling, and the results of the alloy's ability to deform are presented in the form of Schmid and Taylor maps (**Figs. 22 a – h**). These maps demonstrate that, despite a certain degree of deformation, the alloy remains capable of further straining and plastic deformation. The Schmid map describes the potential for activation of slip systems in a crystal under mechanical stress. It utilizes the so-called Schmid factor (m), which depends on the angle between the slip direction (i.e., Burgers vector) and the tangential component of the loading direction at the slip plane. According to the Schmid criterion, slip initiates when the resolved shear stress in the slip direction exceeds a critical value, known as the critical resolved shear stress. This approach helps identify which slip system will be activated first. Schmid maps or diagrams often display isolines of the Schmid factor as a function of the crystal orientation with respect to the loading direction. In contrast, the Taylor map is based on the assumption that multiple slip systems are activated simultaneously to maintain deformation compatibility in a polycrystalline material. The Taylor model illustrates combinations of active slip systems that either minimize the total plastic work or satisfy specific deformation constraints. It is particularly suitable for predicting the macroscopic response of a crystalline material to applied loading. Such maps often display the distribution of strain or stress among individual grains. The color scale on the maps represents the Schmid factor (m) in the case of Schmid maps, and the Taylor factor (M) in the case of Taylor maps. Red areas indicate a high Schmid factor, implying a higher likelihood of slip. Blue areas correspond to a low Schmid factor, indicating a lower probability of slip. This allows identification of crystal (or grain) orientations with the highest tendency for plastic deformation. For Taylor maps, the scale means: blue = low Taylor factor and deformation is easier; red = high Taylor factor and deformation is more difficult (requiring activation of more slip systems). Regions with a high M often behave as harder grains (for example, they deform less readily, which can lead to localized stresses or cracks). As evident from the maps (**Figs. 22 a – d**), during rolling, the Schmid maps were predominantly orange to red, indicating that the alloy has a tendency for plastic deformation. This is consistent with the Taylor maps, which show more green areas, signifying easier, yet not ideal (which would correspond to blue), deformation in this case. For both types of maps, the most typical slip system for BCC crystals was always selected, specifically the {110}<111> system. Both types of maps further indicate that, with increasing reduction during rolling, the size of the regions where slip and plastic deformation occur easily also increases (red and orange areas -Schmid). The black areas on the map correspond solely to poorly indexed regions, not to cracks formed during the rolling process. However, it can be inferred that the grains in the vicinity of the cracks, as well as at the crack sites themselves, were not optimally oriented.

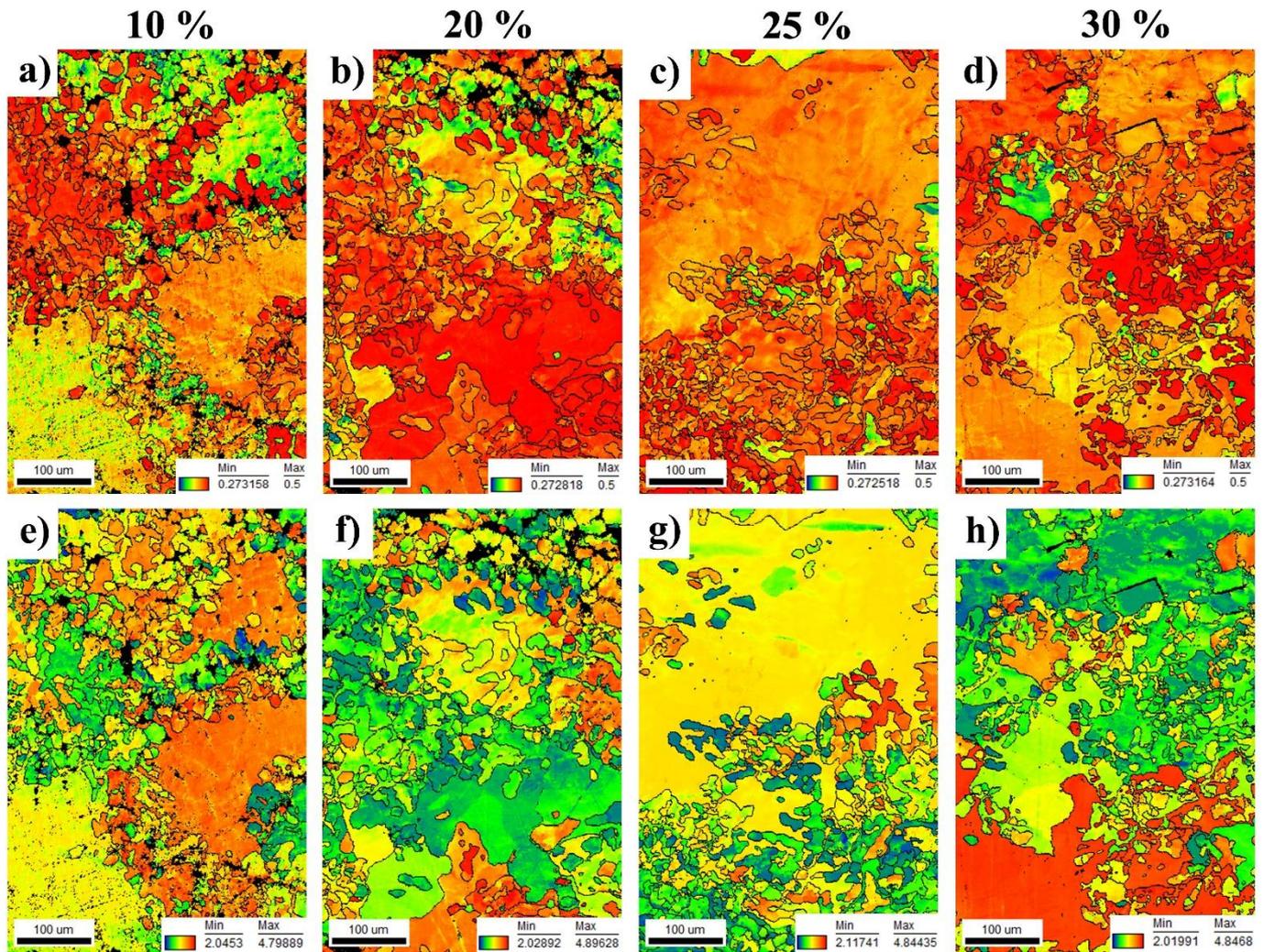

**Fig. 22:** Schmid (a – d) and Taylor maps (e – h) for the annealed cold-rolled MoNbTaTiZr alloy at rolling reduction: a), e) 10 %, b), f) 20 %, c), g) 25 %, d), h) 30 %

During TEM analysis, particles exhibiting distinct striations were observed (**Fig. 16**). The striations may have two possible origins. The first involves the formation of striped particles due to spinodal decomposition, which is typical for BCC systems containing refractory elements [41, 47]. In this process, phase separation occurs at certain temperatures, resulting in a striped or lamellar structure. However, this scenario can be excluded in our case because the particles were found in a deformed sample that was not subjected to any heating or subsequent cooling. The second possible origin of the stripes may indicate the presence of twins or slip bands formed during rolling. If the stripes were caused by precipitated particles present in the structure prior to deformation, these particles would likely exhibit a different crystal structure or orientation compared to the surrounding matrix, and deformation could induce the observed striations within them. However, the particles exhibited a BCC crystal structure and formed only during deformation. Therefore, the observed features correspond to twins within the particles. These twins were identified by TEM as (110) twins (**Fig. 23**). The formation of twins in the CCA alloy based on refractory elements is not surprising, considering the work of Čížek et al. [6], who observed twins in the single-phase BCC HfNbTaTiZr alloy after the early stages of HPT (High Pressure Torsion). The presence of twinning could be beneficial, as it has been shown in previous study [16] that the activation of deformation twinning contributes to improved ductility in the alloy. In the present study, we report another refractory-element-based system with a BCC final structure in which twinning has been observed, in contrast to the well-studied pioneering Nb-Mo-Ta-W and V-Nb-Mo-Ta-W based systems. An increased number of particles exhibiting twinning could therefore represent a potential strategy to mitigate the extreme brittleness of these alloys [1].

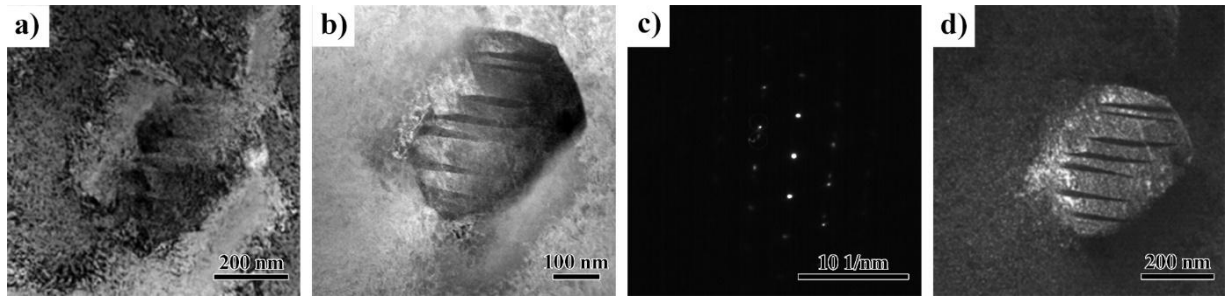

**Fig. 23:** a) TEM image of the twinning in the particle with striations, b) STEM, c) SAED, d) DF

Rolling thus induced the formation of twins within the microstructure, which could have a beneficial effect on the ductility of the alloy. Moreover, a decrease in the Young's modulus in the direction perpendicular to the RD after cold rolling was observed, as shown in **Table 4**. This fact is significant for potential applications in the field of biomaterials, since the value is very close to that of the commonly used Ti-6Al-4V alloy. The high Young's modulus of this alloy was one of the main obstacles limiting its unsuitable as a biomaterial. The anisotropy introduced by rolling can be exploited. **Fig. 24 a** shows the load-displacement curves for the annealed as-cast sample and the sample cold-rolled with a 30 % thickness reduction. From **Fig. 24 a**, it can be observed that there are divergences in the maximum indentation depth, and the slope of the unloading curve determines the hardness of the indented materials. The bright regions enriched with Ta and Mo exhibit higher material stiffness, which is attributed to the presence of Mo (Young's modulus = ~329 GPa). Conversely, regions rich in Ti, Zr, and Nb exhibit a greater deflection of the curve, indicating a decrease in the modulus, which corresponds to the fact that these elements have a significantly lower Young's modulus (Ti = ~105 GPa, Zr = ~88 GPa, and Nb = ~105 GPa). The possible causes of the slight decrease in Young's modulus in the rolled alloy are as follows: Rolling induces the alignment of crystals toward the rolling plane, resulting in a preferential orientation of planes with a lower Young's modulus. Consequently, the average modulus decreases. Rolling may also introduce small defects, which hinder the continuity of the material for elastic deformation. The load–displacement curves (**Fig. 24 a**) do not indicate the presence of the phenomenon known as pop-in. This may indicate that the alloy does not exhibit, or more likely does not clearly show, the pop-in phenomenon. This could be related to the polishing conditions, as the pop-in effect is highly sensitive to the surface condition. If the surface has been mechanically polished, the pop-in load may be significantly reduced or even completely suppressed. The observed wavy profile of the curve is likely associated with the repeated occurrence of a mechanism in which the slip motion of dislocations is temporarily impeded by obstacles within the crystal lattice and subsequently resumes upon overcoming these barriers. The consistently lower slope observed in the dark phase implies a reduced work-hardening ratio and, consequently, greater deformability compared to the bright phase. To assess the degree of work hardening, Load/Displacement versus Displacement curves were plotted (**Fig. 24 b**). The slope of these curves (taken from the loading segment of the curve) provides an estimate of the work-hardening ratio. The slope of the load–displacement curve is regarded as an indicator of the degree of work hardening. Based on the slope, it is evident that the bright phases exhibit a higher extent of work hardening during deformation compared to the dark phase (**Fig. 24 b**). This is closely related to the chemical composition of the respective regions.

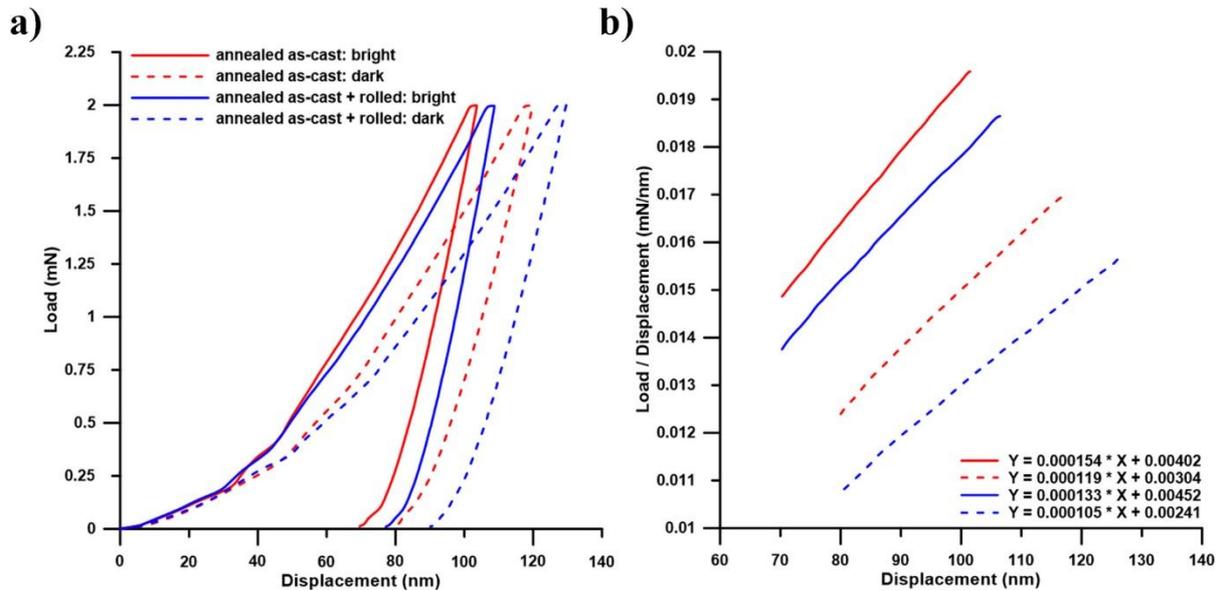

**Fig. 24:** a) Representative indentation load-displacement curves for a loading of 2 mN, b) work hardening ratio determination plot

## Conclusions

In summary, this pilot study provides a comprehensive characterization of the MoNbTaTiZr complex concentrated alloy in its as-cast, homogenized, and cold-rolled conditions. The results demonstrate that homogenization annealing refines the microstructure by promoting phase growth and reducing grain boundary segregation through the formation of Ti-Zr-based particles. Cold rolling leads to significant grain fragmentation and the emergence of a secondary FCC phase identified as the $Zr_2Ta$ Laves phase. Although distinct rolling textures were not observed, rotational relationships between texture components were identified, indicating complex orientation relationships during deformation. Schmid and Taylor factor analyses confirm the alloy still retains the capacity for plastic deformation, and the observation of deformation twinning suggests an additional mechanism for enhanced ductility. These findings highlight the potential of this refractory-based complex concentrated alloy for applications requiring a balance of strength and ductility under severe plastic deformation.

## Acknowledgement

The authors acknowledge the assistance provided by the Ferroic Multifunctionalities project, supported by the Ministry of Education, Youth, and Sports of the Czech Republic. Project No. CZ.02.01.01/00/22_008/0004591, co-funded by the European Union. CzechNanoLab project LM2023051 funded by MEYS CR is gratefully acknowledged for the financial support of the measurements/sample fabrication at LNSM Research Infrastructure. The authors would also like to thanks to Jiří Zýka and Tomáš Chmela (both from UJP PRAHA a.s.) for casting the refractory complex concentrated alloys.

## Declaration of competing interest

The authors declare no conflict of interest.

## Data availability

The used data are accessible *via* the Zenodo repository: https://doi.org/10.5281/zenodo.16738117.